\documentclass[aps,prb,twocolumn,groupedaddress,showpacs,superscriptaddress]{revtex4-1}
\usepackage{amsmath}
\usepackage{graphicx}
\usepackage{verbatim}
\usepackage[caption=false]{subfig}
\usepackage[bookmarks=false,colorlinks=true, linkcolor=blue, citecolor=blue, urlcolor=blue]{hyperref}

\begin{document}

\title{Meissner transmon qubit - architecture and characterization}

\author{Jaseung Ku}
\affiliation{Department of Physics, University of Illinois at Urbana-Champaign, Urbana, Illinois 61801, USA}

\author{Zack Yoscovits}
\affiliation{Department of Physics, University of Illinois at Urbana-Champaign, Urbana, Illinois 61801, USA}

\author{Alex Levchenko}
\affiliation{Department of Physics and Astronomy, Michigan State University, East Lansing, Michigan 48824, USA}

\author{James Eckstein}
\affiliation{Department of Physics, University of Illinois at Urbana-Champaign, Urbana, Illinois 61801, USA}

\author{Alexey Bezryadin }
\affiliation{Department of Physics, University of Illinois at Urbana-Champaign, Urbana, Illinois 61801, USA}

\date{\today}

\begin{abstract}
We present a new type of transmon split-junction qubit which can be tuned by Meissner screening currents in the adjacent superconducting film electrodes. The best detected relaxation time ($T_1$) was of the order of 50 $\mu$s and the dephasing time ($T_2$) about 40 $\mu$s. The achieved period of oscillation with magnetic field was much smaller than in usual SQUID-based transmon qubits, thus a strong effective field amplification has been realized. This Meissner qubit allows an efficient coupling to superconducting vortices. We present a quantitative analysis of the radiation-free energy relaxation in qubits coupled to Abrikosov vortices. The observation of coherent quantum oscillations provides strong evidence that vortices can exist in coherent quantum superpositions of different position states. According to our suggested model, the wave function collapse is defined by Caldeira-Leggett dissipation associated with viscous motion of the vortex cores. 
\end{abstract}

\pacs{03.67.Lx, 74.50.+r, 85.25.Am}

\maketitle

\section{Introduction}\label{sec:intro}
Several promising architectures for quantum computers are based on the superconducting qubits~\cite{Makhlin-RMP01,Devoret-Martinis,Nori-PT05} based on Josephson junctions (JJs). These designs utilize either charge~\cite{Nakamura-Nature99}, phase~\cite{Martinis-PRL02} or flux~\cite{Friedman-Nature00,Mooij-Science00} degrees of freedom. These systems have made tremendous progress in recent years in realizing increasingly sophisticated quantum states, measurements and operations with high fidelity~\cite{Devoret-Review}. Superconducting quits are also attractive technologically because they can be naturally integrated into large-scale quantum circuits~\cite{BarendsPRL2013,Schoelkopf-Girvin}. 
However, this main advantage of superconducting qubits brings a substantial challenge at the same time since strong coupling also implies a substantial interaction between the qubits and their environment, which can break quantum coherence. Understanding the limiting factors of qubit operation is of fundamental and practical importance. Previously, a separation of various contributing factors to the qubit relaxation and decoherence has been achieved~\cite{Astafiev-PRB04,Bertet-PRL05,Martinis-PRL05,Neeley-PRB08}. Abrikosov vortex is one important example of such environment. Dissipation caused by vortices has also been studied in a  superconducting resonator~\cite{SongPRB2009}. Recently, vortices have been shown to trap quasiparticles in superconducting resonator and qubit, leading to the increase in the quality factor of resonator~\cite{NsanzinezaPRL2014} and the relaxation time of qubit~\cite{wang_measurement_2014}.

One interesting theoretical possibility of drastically improving quantum coherence in qubits is to couple them to Majorana fermions~\cite{beenakker2013}. A qubit based on Majorana states is expected to exhibit especially long coherence times. One approach to create Majorana states is to deposit a superconductor onto a topological insulator and to create vortices in the superconductor. In this case Majorana states can nucleate in the vortex core~\cite{beenakker2013}. Thus a study of qubits coupled to vortices is needed, in order to determine whether a qubit coupled to a vortex can preserve its quantum coherence and for how long. Here we use Meisser qubits to quantify dissipation effects produced by single vortices. 

A major advance in the superconducting qubit performance became possible after the invention of the \textit{transmon qubit}~\cite{koch2007PRA}. When combined with the three-dimensional circuit-quantum electrodynamics (cQED) platform developed in Ref.~\cite{Paik2011PRL}, the transmon has shown huge improvements of the relaxation time, up to several hundreds microseconds~\cite{Rigetti2012PRB,Oliver2013MRS}. Like the common transmon our device involves a capacitance linked by a nonlinear kinetic inductance (JJs). The main difference is that our qubit is coupled to the Meissner current and the supercurrents generated by vortices. Yet the relaxation time of such device, designed to probe the environment is rather large, namely about 50 $\mu$s in the best case. We argue that the limiting factor was the Purcell effect, thus the relaxation time can be made even longer if necessary.

Because in our qubit design the Meissner current is allowed to flow, partially, into the qubit, a significant amplification of the magnetic field effect is demonstrated. The qubit transition frequency is periodically modulated by the applied magnetic field, but the period is much smaller compared to the value estimated by dividing the flux quantum by the qubit loop geometric area. 

The Meissner qubit allows a strong coupling to the vortices in the leads. We perform a detailed study of the radiation-free decoherence effects produced by the vortex cores. It should be stressed  that the qubit relaxation time may be shortened by the presence of a vortex in the superconducting film, due to the Bardeen-Stephan viscous vortex flow. No quantitative study has been done so far to test how qubit quantum states relax due to coupling to Abrikosov vortices. Our key finding is that vortices can remain, over many microseconds, in quantum superposition states generated because the Lorentz force experiences strong quantum fluctuations. The relaxation rate added to the qubit by each single vortex was measured and appears to be surprisingly low, of the order of 10 kHz. We propose a semi-quantitative model which allows us to estimate this radiation-free relaxation rate caused by viscous flow of vortices. Up until now it was well established that classical supercurrents can generate heat through viscous flow of vortices~\cite{SongPRB2009}. Now we establish that quantum superposition currents, such as those existing in qubits and characterized by zero expectation value, can also generate heat through the same mechanism. Such heat dissipation occurs through the spread of the wavefunction of the vortex center followed by a collapse of this smeared wave function.  

\begin{figure}[t]
\begin{subfloat}{\label{fig:setup_a}}
\end{subfloat}
\begin{subfloat}{\label{fig:setup_b}}
\end{subfloat}
\begin{subfloat}{\label{fig:setup_c}}
\end{subfloat}
\begin{subfloat}{\label{fig:setup_d}}
\end{subfloat}
\centering
\includegraphics[width=0.98\columnwidth]{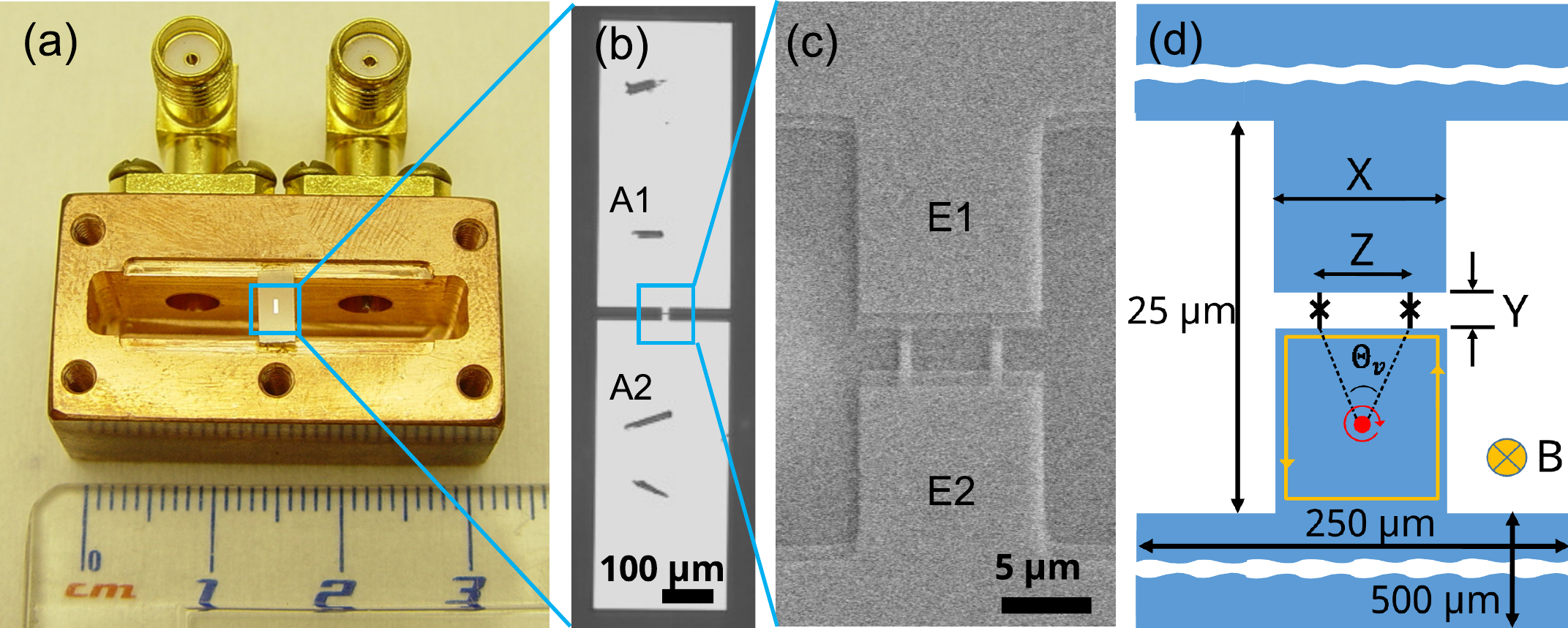}
\caption{(a) Optical image of the Meissner transmon qubit fabricated on a sapphire chip, which is mounted in the copper cavity. (b) A zoomed-in optical image of the qubit. Two rectangular pads marked A1 and A2 act as an RF antenna  and shunt capacitor. (c) Scanning electron microscope (SEM) image of the electrodes marked E1 and E2, and a pair of JJs. (d) Schematics of the Meissner qubit. The X, Y and Z denote the width, the distance between the electrodes, and the distance between two JJs, which are indicated by $\times$ symbols. The red dot and circular arrow around it in the bottom electrode represent a vortex and vortex current flowing clockwise, respectively. $\Theta_\text{v}$ is a polar angle defined by two dashed lines connecting the vortex and two JJs. The orange rectangular loop on the boundary of the bottom electrode indicates the Meissner current circulating counterclockwise. }\label{fig:setup}.  
\end{figure}

\section{Experimental Results}\label{sec:results}
\subsection{Qubit frequency modulation}
\begin{figure}[t]
\begin{subfloat}{\label{fig:TransmVsB}}
\includegraphics[width=\columnwidth]{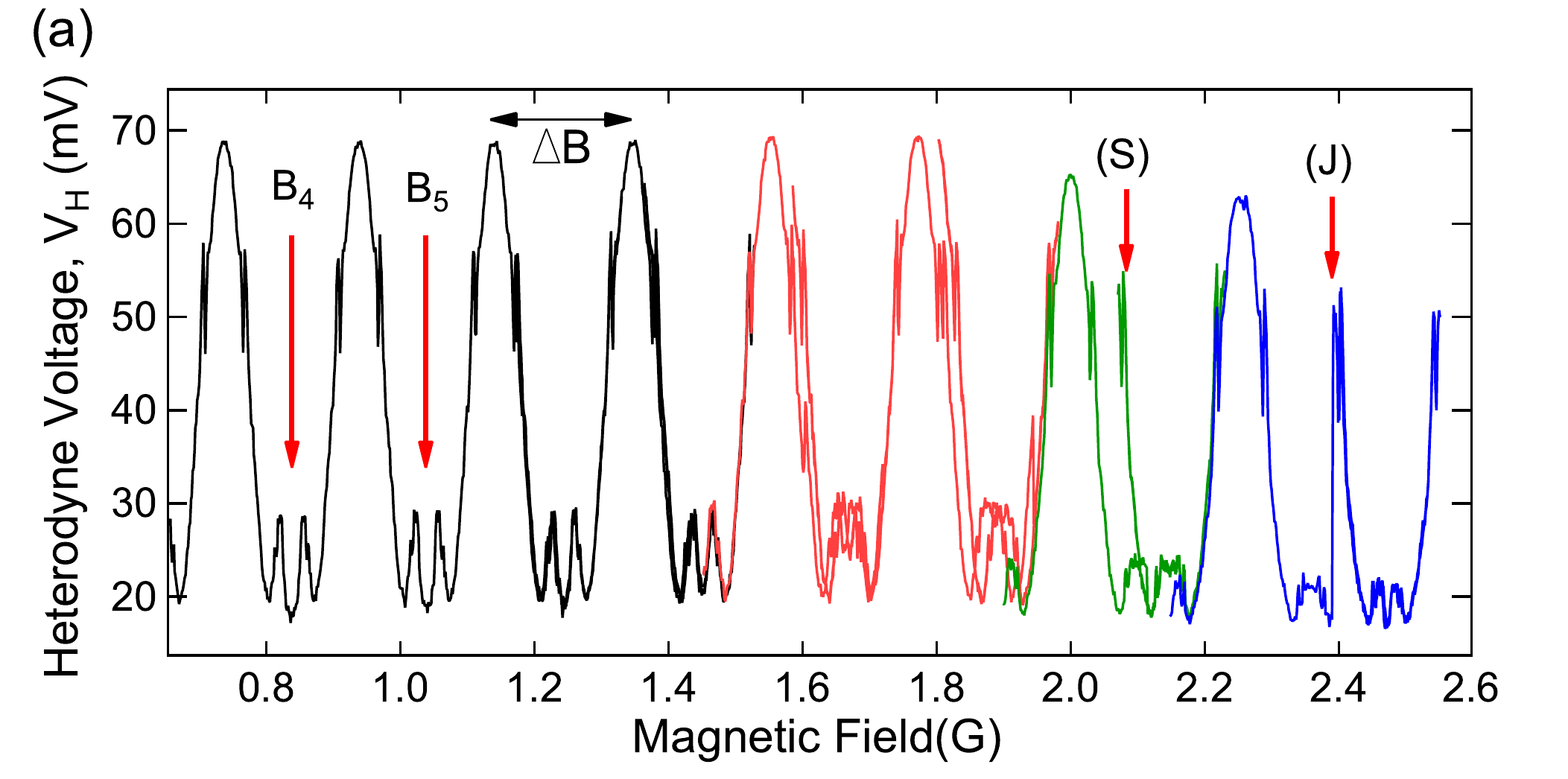}
\end{subfloat}
\begin{subfloat}{\label{fig:EffArea}}
\includegraphics[width=\columnwidth]{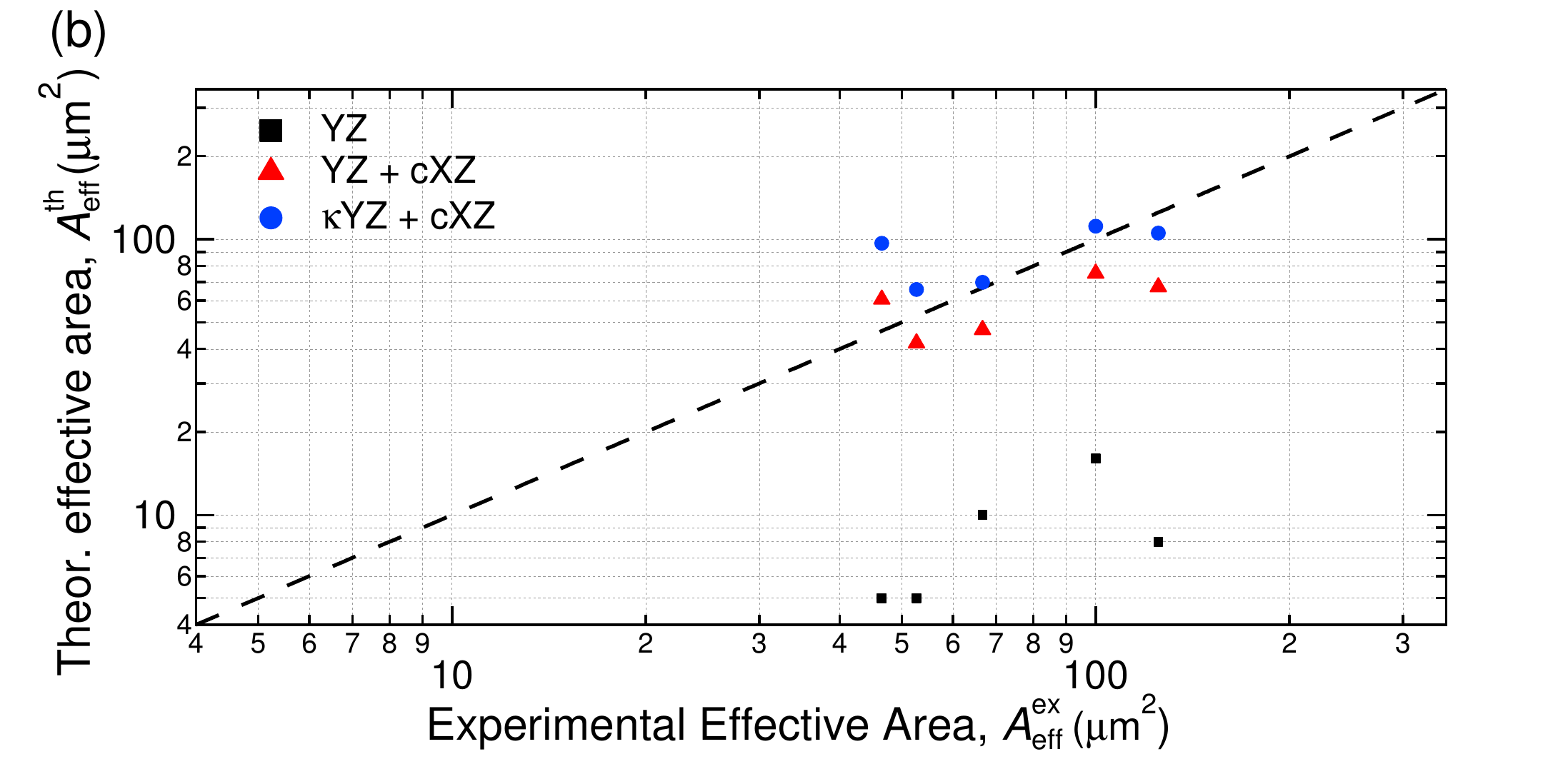}
\end{subfloat}
\caption{\label{fig:Bdep}(Color online) (a) Periodic heterodyne voltage oscillation (``HV-oscillation'') as a function of magnetic field. Four different colors represent the separate measurement runs where the magnetic field swept either round-trip or one-way. $\Delta B$ shows the period of the modulation. The arrows marked $B_4$ and $B_5$ indicate the positions of two adjacent sweet spots, characterized by $df_{01}/dB$=0. Each sweet spot is equivalent to $B$=0 state. The actual B=0 sweet spot is outside the range of the plot. The mark ``S'' indicates the field at which the periodic signal was shifted to the right, revealing some hysteresis caused by vortex entrance to the electrodes. This hysteretic behavior can be seen in the second (red) segment of the B-field sweep ranging from about 1.45 to 2 Gauss. The mark ``J'' shows the moment when a vortex (or a small group of vortices) entered the electrodes during the forward-sweeping magnetic field. (b) Comparison of theoretical ($A^\text{\tiny th}_\text{\tiny eff}$) and experimental ($A^\text{\tiny ex}_\text{\tiny eff}$) effective areas of five samples. The dashed line depicts the ideal case of $A^\text{\tiny th}_\text{\tiny eff}$=$A^\text{\tiny ex}_\text{\tiny eff}$. See text for the definition of these quantities.}
\end{figure}

\begin{figure}[t]
\begin{subfloat}{\label{fig:spec}}
\includegraphics[width=0.7\columnwidth]{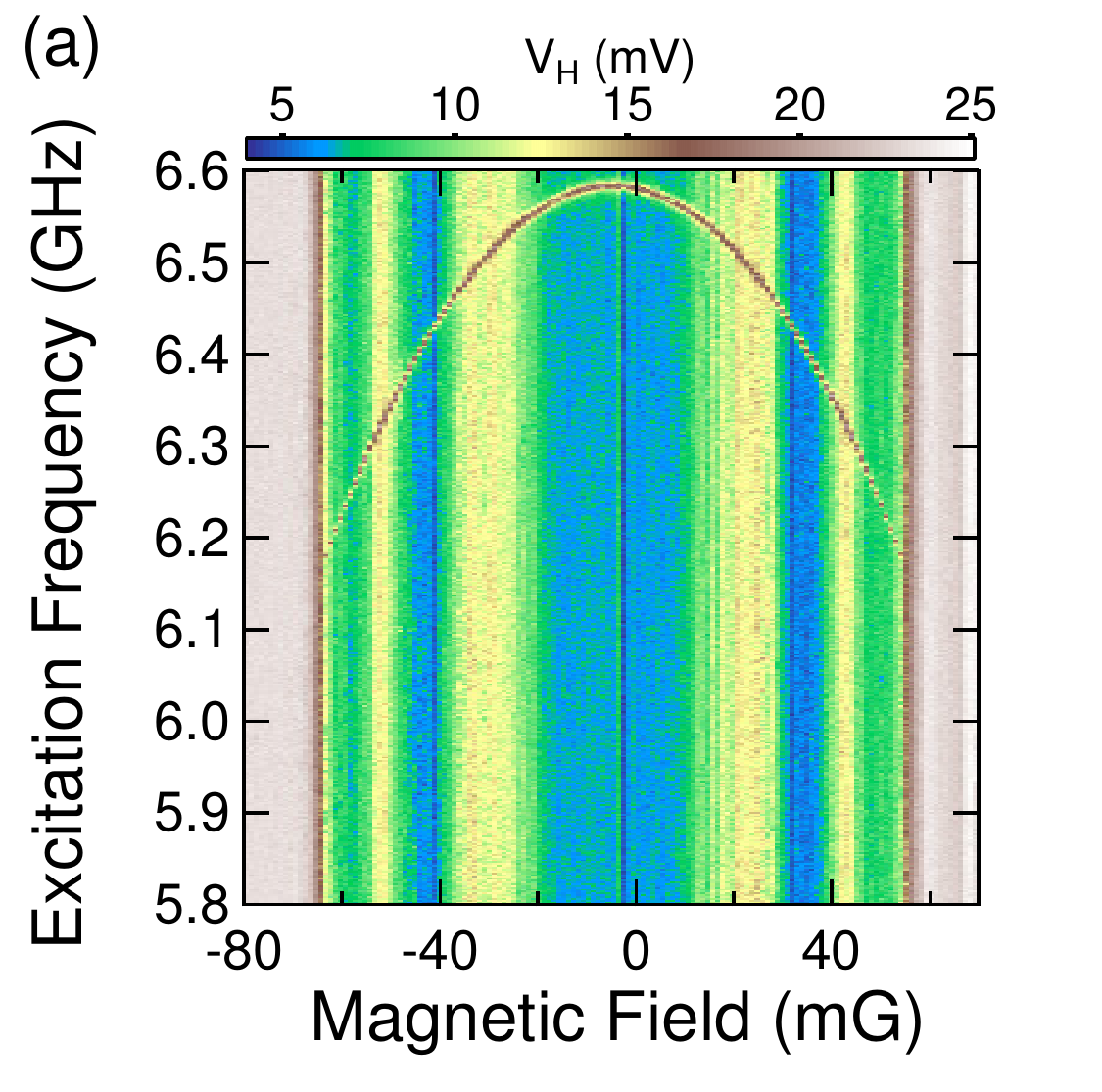}
\end{subfloat}
\begin{subfloat}{\label{fig:specfit}}
\includegraphics[width=0.7\columnwidth]{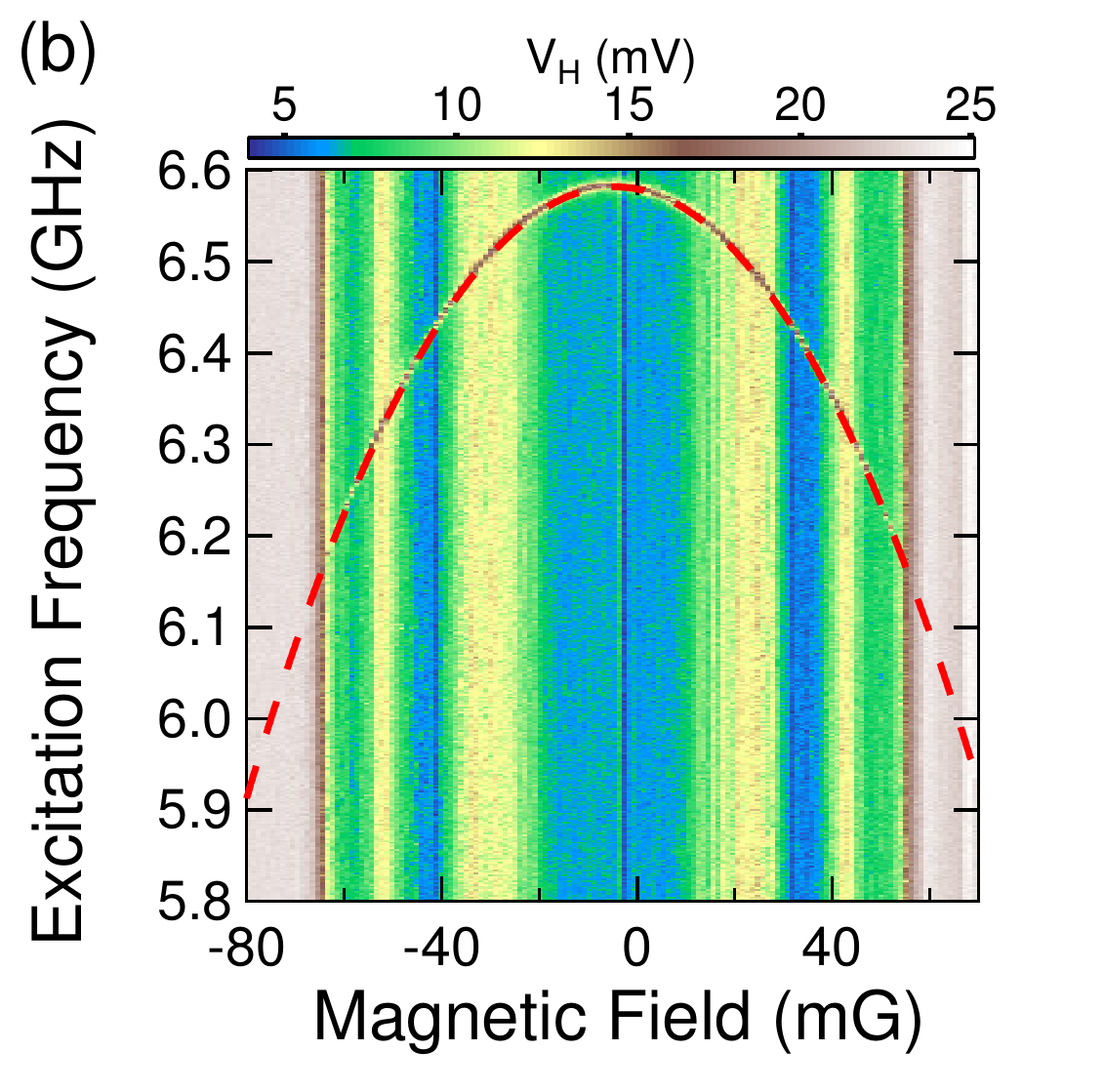}
\end{subfloat}
\caption{\label{fig:Spec}(Color online) (a) Spectroscopy of Meissner transmon (N1) as a function of applied magnetic field. This is raw data.  (b) The parabola-like dashed line shows a phenomenological fit to the qubit transition frequency $f_{01}$ versus magnetic field $B$.}
\end{figure}

The design of our devices is shown in Fig.~\hyperref[fig:setup]{\ref{fig:setup}}, while details of the fabrication and measurement techniques are described at length in the Appendix~\hyperref[sec:device]{\ref{sec:device}}. In brief, the qubits have been placed inside a three-dimensional (3D) microwave cavity made of Cu (Fig.~1a). The state of the qubit has been determined by measuring the transmission of the 3D cavity.  First, we investigated the magnetic field dependence of the qubits, anticipating to observe periodic SQUID-type oscillations. The transmission versus the magnetic field (``B-field'') varied as shown in Fig.~\hyperref[fig:TransmVsB]{\ref{fig:TransmVsB}}. This plot represents the heterodyne voltage, produced by mixing a microwave signal passing through the cavity containing the qubit and a reference signal. During this measurement the qubit remains in its ground state. Yet the cavity input power is chosen such that the transmission of the cavity is the most sensitive to the qubit transition from the ground to the excited state, i.e., the maximum-contrast power was used [\onlinecite{Reed2010PRL}]. The probing microwave frequency for this measurement equals the bare cavity frequency. Four segments in different color (color online) represent four separate measurement runs. The magnetic field was swept round-trip (up and down) in the first three segments (black, red and green), while it was swept one-way (up) in the last segment (blue). The modulation of the transmission at low $B$-field arises from the change of the onset power---the lowest power at which the cavity starts to show a sharp increase in transmission, which was termed ``bright state'' (near-unity transmission) in Ref.~\onlinecite{Reed2010PRL}. The onset power depends on the difference between the qubit transition frequency ($f_{01}$) and the bare cavity frequency ($f_\text{c}$). The key point is that $f_{01}$ is modulated periodically by the applied magnetic field, because the qubit includes a SQUID-like loop formed by the two JJs and the two electrodes, marked in Fig.1c as E1 and E2. The heterodyne voltage $V_\text{H}$ is proportional to the microwave transmission. Thus, as magnetic field was increased, we observed periodic or quasi-periodic heterodyne voltage oscillation (``HV-oscillation''). The voltage changes reproducibly and periodically with magnetic field, up to the first critical field of the electrodes, $B_{c1}\approx$1.6 Gauss. This is the critical field at which Abrikosov vortices begin to enter the electrodes. The period of the HV-oscillation, $\Delta B$, can be defined as the distance between the adjacent principle maxima, as is illustrated in Fig.~\hyperref[fig:Bdep]{\ref{fig:TransmVsB}} by the horizontal arrow. Equivalently, the period can be defined as the separation between the so-called \textit{sweet spots}. Some of these sweet spots, namely $B_4$ and $B_5$ are indicated by the vertical arrows in Fig.~\hyperref[fig:Bdep]{\ref{fig:TransmVsB}}. The sweet spots are the points equivalent to zero magnetic field. It should be reminded here that if the device is tuned to a sweet spot then it is insensitive, in the first order, to the flux noise, because $df_{01}/dB$=0 [see Fig.~\hyperref[fig:spec]{\ref{fig:spec}}] and $dV_\text{H}/dB$=0 [see Fig.~\hyperref[fig:Bdep]{\ref{fig:TransmVsB}}].

At sufficiently low fields, when there are now vortices in the electrodes, the sweet spots occur periodically because the critical current of the SQUID loop changes periodically with magnetic field.
In our case the design is such that the phase gradient created by vortices present in the electrodes couples to the SQUID loop. Therefore the exact periodicity of the sweet spots becomes broken when vortices begin to enter the electrodes at $B>B_{c1}$. Thus our device acts as a vortex detector. Entrapment of vortices inside the electrodes makes the transmission hysteretic with magnetic field up-down sweeps. An example of such hysteresis is clearly seen at the position marked ``S'' in Fig.~\hyperref[fig:TransmVsB]{\ref{fig:TransmVsB}}. In addition to the period increases and the hysteresis, we also observe abrupt jumps in the transmission. One such jump is marked ``J'' in Fig.~\hyperref[fig:TransmVsB]{\ref{fig:TransmVsB}}. The jumps indicate that the vortex enters in the near proximity of the qubit loop and thus should be strongly coupled to the qubit state. 

\begin{table}[t] 
\caption{\label{tab:table1} Comparison of the experimental and theoretical periods for five samples. $\Delta B$ is the measured period of oscillation. $\Delta B_{\text{\tiny YZ}}$, $\Delta B_{\text{\tiny YZ+cXZ}}$, and $\Delta B_{\text{\tiny kYZ+cXZ}}$ are the theoretical periods calculated using three different effective areas denoted by the subscripts.}
 \begin{ruledtabular}
 \begin{tabular}{lcccllll}
Sample & X &Y & Z & $\Delta B$& $\Delta B_{\text{\tiny YZ}}$ & $\Delta B_{\text{\tiny YZ+cXZ}}$ & $\Delta B_{\text{\tiny kYZ+cXZ}}$\\
         & ($\mu$m) &($\mu$m) &($\mu$m) & (G) & (G) & (G) & (G) \\
\hline
N1 & 10 & 1 & 5 & 0.38 & 4 & 0.49 & 0.30\\
N2 & 10 & 2 & 5 & 0.3  & 2 & 0.43 & 0.29\\
N3 & 15 & 1 & 5 & 0.43  & 4 & 0.34 & 0.22\\
N6 & 10 & 2 & 8 & 0.16  & 2.5 & 0.30 & 0.19\\
N7 & 10 & 2 & 8 & 0.2  & 1.25 & 0.27 &0.18 \\
 \end{tabular}
 \end{ruledtabular}
 \end{table}

In our qubits, the effective Josephson energy is modulated by the external magnetic field, but the period is set differently as compared to the typical SQUID-type device~\cite{Schreier2008PRB}. Unlike in usual SQUIDs, the modulation of the critical current in the present case is driven mostly by the Meissner currents in the electrodes, and to a much lesser extent by the magnetic flux through the SQUID loop. This can be seen from the fact (See Table~\hyperref[tab:table1]{\ref{tab:table1}}) that the experimental period, $\Delta B$, is much smaller than the period computed using the area of the superconducting loop, $\Delta B_{\text{\tiny YZ}}$.
This is why the sensitivity of the Meissner qubit energy to the external field is higher compared to the usual split-junction transmon~\cite{Schreier2008PRB}.  

One can understand this new period by considering the phase constraint~\cite{Hopkins2005science,Pekker2005PRB}:
\begin{equation}\label{eq:phaseconstraint}
 \theta_1-\theta_2+2\delta (B)=2\pi n_\text{v},
\end{equation}
where $\theta_{1,2}$ is the phase difference across each JJ, $\delta (B)$ the phase difference generated in the thin-film electrodes by the Meissner current and defined as the phase difference between the entrance points of of the JJ bridges, i.e., between the bridges in which JJs are created. Here $n_\text{v}$ is the vorticity. Since the current-phase relationship of JJs is single-valued, and the inductance of the JJ bridges and the electrodes is very small, the vorticity is always zero, $n_\text{v}=0$, in our SQUID-type devices, just like in  common SQUIDs. The field-dependent phase accumulation is $\delta (B)$=$\int{\vec{\nabla}\varphi(B)\cdot d \vec{l}}$, where $\nabla\varphi (B)$ is the phase gradient of the order parameter in the electrodes. The gradient originates from the Meissner (screening) current in the \textit{electrodes}, if there are no vortices. At sufficiently high fields, at which vortices enter the electrodes, an additional contribution to the total phase gradient occurs due to the vortices.  Following the Ref.~\onlinecite{Pekker2005PRB}, the magnetic period can be estimated as
\begin{equation}\label{eq:period}
\Delta B = \left[ \left(\frac{\Phi_0}{cXZ}\right)^{-1}+\left(\frac{\Phi_0}{YZ}\right)^{-1}\right]^{-1},
\end{equation}
where the numerical coefficient $c=(8/\pi^2)\sum^{\infty}_{n=0}(-1)^n/(2n+1)^2\approx 0.74$ can be found by solving appropriate boundary problem for the Laplace equation. We cautiously notice that the Eq.~\eqref{eq:period} may not be strictly applicable to our case because it was derived for mesoscopic electrodes, where $\lambda_\perp$ is much larger than $X$, the width of the electrode. In our samples $\lambda_\perp\approx550$ nm $<X=10\text{~or~ } 15$ $\mu$m. Meissner currents are stronger in the case of a relatively small perpendicular magnetic length, therefore this model may still provide a semi-quantitative estimate. To achieve a satisfactory agreement we will have to introduce corrections related to the field focusing effect.

We emphasize that unlike a regular SQUID the period is not set by the area $YZ$ enclosed by SQUID loop only, but rather by a much larger effective area $YZ+cXZ$; the dimensions re indicated in Fig.~\hyperref[fig:setup]{\ref{fig:setup_d}}~\cite{Hopkins2005science, Pekker2005PRB}. To compare those two effective areas, we plot in Fig.~\hyperref[fig:EffArea]{\ref{fig:EffArea}} the theoretical effective areas $A^{\text{\tiny th}}_\text{\tiny eff}$  versus the ``experimental'' effective area $A^{\text{\tiny ex}}_\text{\tiny eff}=\Phi_0/\Delta B$ for five samples. Here $\Delta B$ is the low-field, vortex-free period of the HV-oscillation for each sample. The black squares and the red triangles represent, respectively, the theoretical effective area calculated as $A^{\text{\tiny th}}_\text{\tiny eff}$=$YZ$ (geometric SQUID loop area approach) and as $A^{\text{\tiny th}}_\text{\tiny eff}$=$YZ+cXZ$ (Meissner current phase gradients approach). The dashed curve represents the ideal case, $A^{\text{\tiny th}}_\text{\tiny eff}$=$A^{\text{\tiny ex}}_\text{\tiny eff}$. The red triangles appear much closer to the ideal dashed line. Therefore the qubit energy is controlled mostly by the Meissner currents, which produce a strong phase bias of the SQUID loop.

Another possible contribution to the observed amplification of the magnetic sensitivity is the focusing of the magnetic field into the SQUID loop area by the superconducting electrodes. Such focusing is also due to Meissner effect, which is strong due the relatively small $\lambda_{\perp}$. This field focusing effect enhances the magnetic field by a factor of $\kappa=B_1/B_0 >1$. The ratio of the field $B_1$, enhanced by the field-focusing, to the applied field $B_0$ is estimated in Appendix~\hyperref[sec:kappa_cal]{\ref{sec:kappa_cal}}. Thus the effective area set by the SQUID loop increases by a factor of $\kappa$. To incorporate the field focusing effect, we replaced $YZ$ with $\kappa YZ$ in Eq.~(\ref{eq:period}). The result is plotted in Fig.~\hyperref[fig:Bdep]{\ref{fig:EffArea}}, showing an improved and now quite satisfactory agreement with the experimental results. 

Now, we turn to the magnetic field dependence of the qubit energy. For the spectroscopy, the qubit was excited with 2 $\mu$s long saturation pulse, which was immediately followed by a few microsecond readout pulse. The excitation frequency was swept up---low to high frequency---with a fixed step size at a fixed magnetic field, and this process was repeated for equally spaced magnetic field. Fig.~\hyperref[fig:spec]{\ref{fig:spec}} shows the 2D color plot of the transmission as a function of the excitation frequency and external magnetic fields. The color represents the heterodyne voltage of the transmission of the cavity.  The dashed line is a fit to the qubit spectrum with the following fit function: $f_{01}$=$f_{0}\sqrt{|\cos (\pi (B-B_0)A_\text{\tiny eff}/\Phi_0)|} $, where  $f_0$, $B_0$ and $A_\text{\tiny eff}$ are the fitting parameters. We used the approximate relation for $f_{01}$=$\sqrt{8E_\text{J} E_\text{C}}/h$, where $E_\text{J}$=$\hbar I_\text{c}/2e$, and $I_\text{c}(B)$=$2I_\text{c1}|\cos (\pi\Phi/\Phi_0)|$=$2I_\text{c1}|\cos (\pi BA_\text{\tiny eff}/\Phi_0)|$. $B_0$ and $A_\text{\tiny eff}$ are offset field to account for the residual magnetic field and the effective area, respectively. The best fit values of $B_0$ and $f_0$ were $-4.5$ mG and 6.583 GHz. The best fit effective area was $A_\text{\tiny eff}=55.2$ $\mu$m$^2$, which is consistent with the $A_\text{\tiny eff}=55$ $\mu$m$^2$ determined from the periodic oscillations of Fig.~\hyperref[fig:Bdep]{\ref{fig:TransmVsB}}.

\subsection{Time domain measurement: low magnetic field}
\begin{figure}[t,b]
\begin{subfloat}{\label{fig:Timedomain_T1_inset}}
\includegraphics[width=\columnwidth]{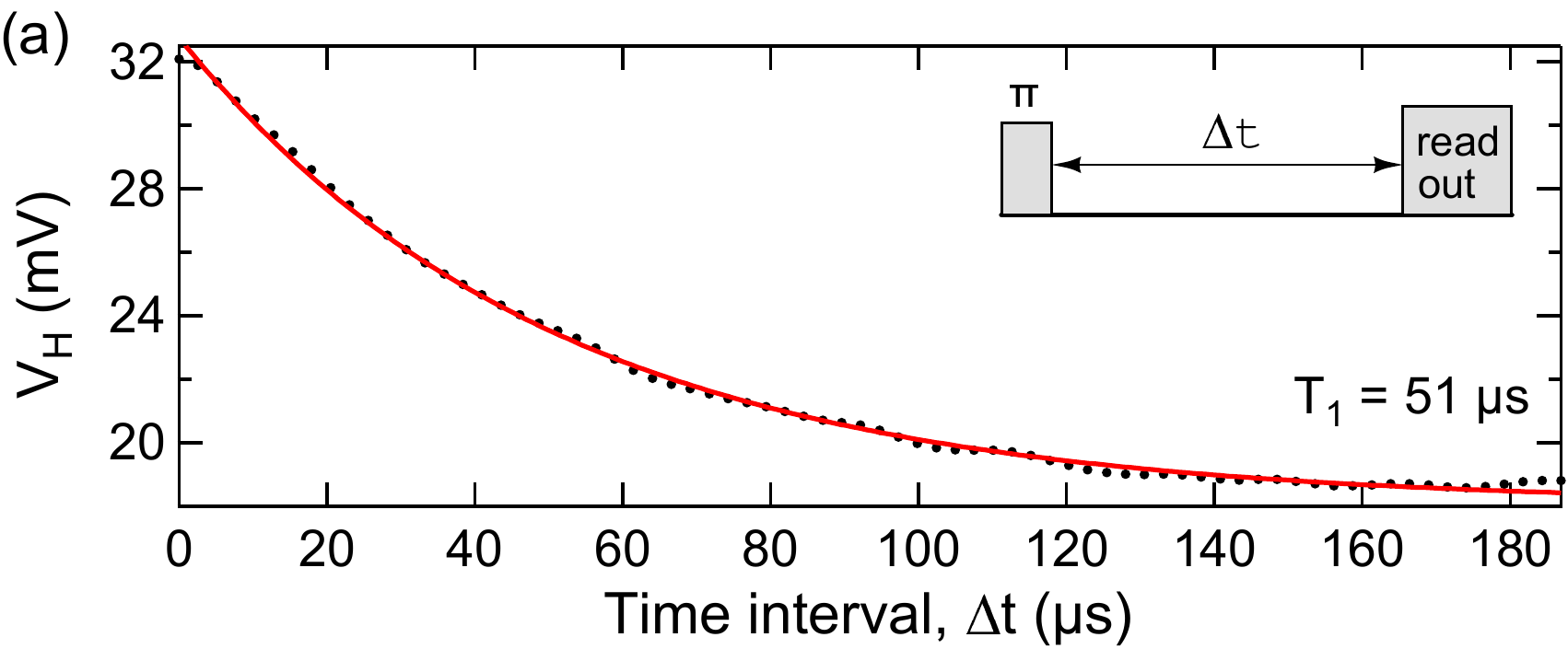}
\end{subfloat}
\begin{subfloat}{\label{fig:Timedomain_ramsey_inset}}
\includegraphics[width=\columnwidth]{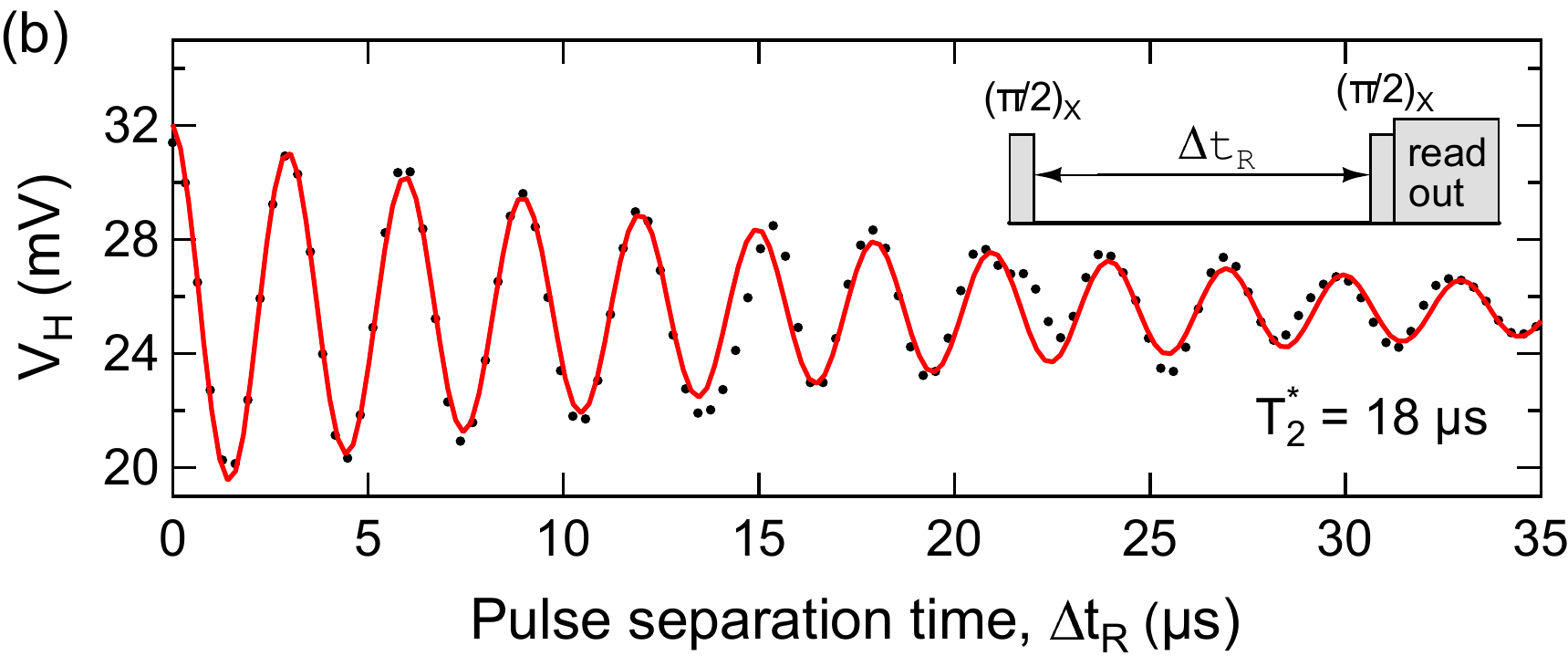}
\end{subfloat}
\begin{subfloat}{\label{fig:Timedomain_echo_inset}}
\includegraphics[width=\columnwidth]{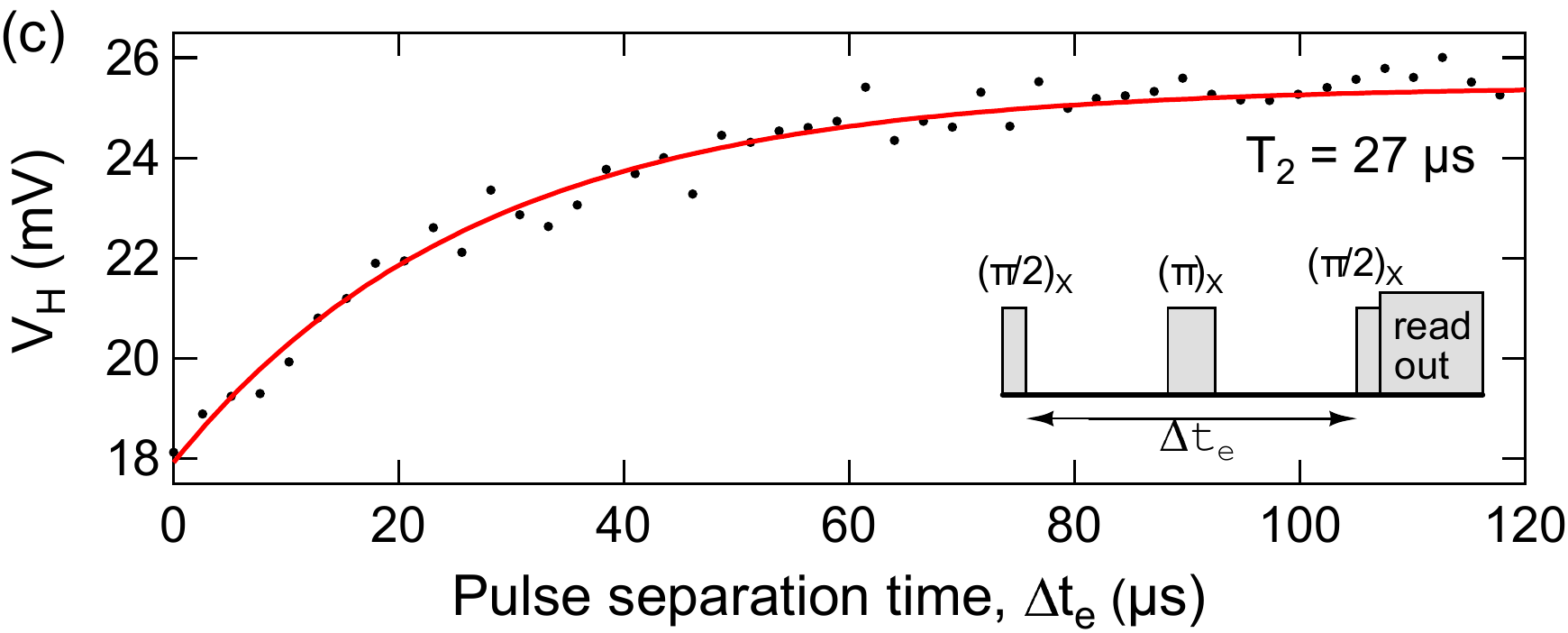}
\end{subfloat}
\caption{\label{fig:Timedomain} (Color online) Time domain measurements of the N7 sample at $B =7.5$ mG. (a) Relaxation time measurement ($T_1=51$ $\mu$s) (b) Ramsey fringe experiment ($T_2^*=18$ $\mu$s)  (c) Hahn spin echo experiment ($T_2=27$ $\mu$s). The red solid lines are the fits to the data. See the main text for the fitting functions.}
\end{figure}

Now we will look into the conventional time-domain measurements under the applied magnetic field. The time-domain measurements shown in Fig.~\hyperref[fig:Timedomain]{\ref{fig:Timedomain}} were performed to measure three time scales: relaxation time ($T_1$), phase coherence time ($T^*_2$) by Ramsey fringe, and phase coherence time ($T_2$) by Hahn spin echo.

For the relaxation time measurement, we applied $\pi$ pulse (100 ns) first and then read out the qubit state after the time interval $\Delta t$. For the Ramsey measurement, we applied two $\pi/2$ pulses (50 ns) separate by pulse separation time $\Delta t_\text{R}$, and then readout was performed immediately after the second $\pi/2$ pulse. Similarly, in spin echo measurement, the measurement protocol was as the Ramsey protocol, except that an additional $\pi$ pulse was inserted right in the middle of the two $\pi/2$ pulses. The separation between the two $\pi/2$ pulses is denoted by $\Delta t_\text{e}$ [see the inset in Fig.~\hyperref[fig:Timedomain]{\ref{fig:Timedomain_echo_inset}}]. 

\begin{figure}[t]
\makebox[\linewidth][c]{
\begin{subfloat}{\label{fig:lowB_N1}}
\centering
\includegraphics[width=0.5\columnwidth]{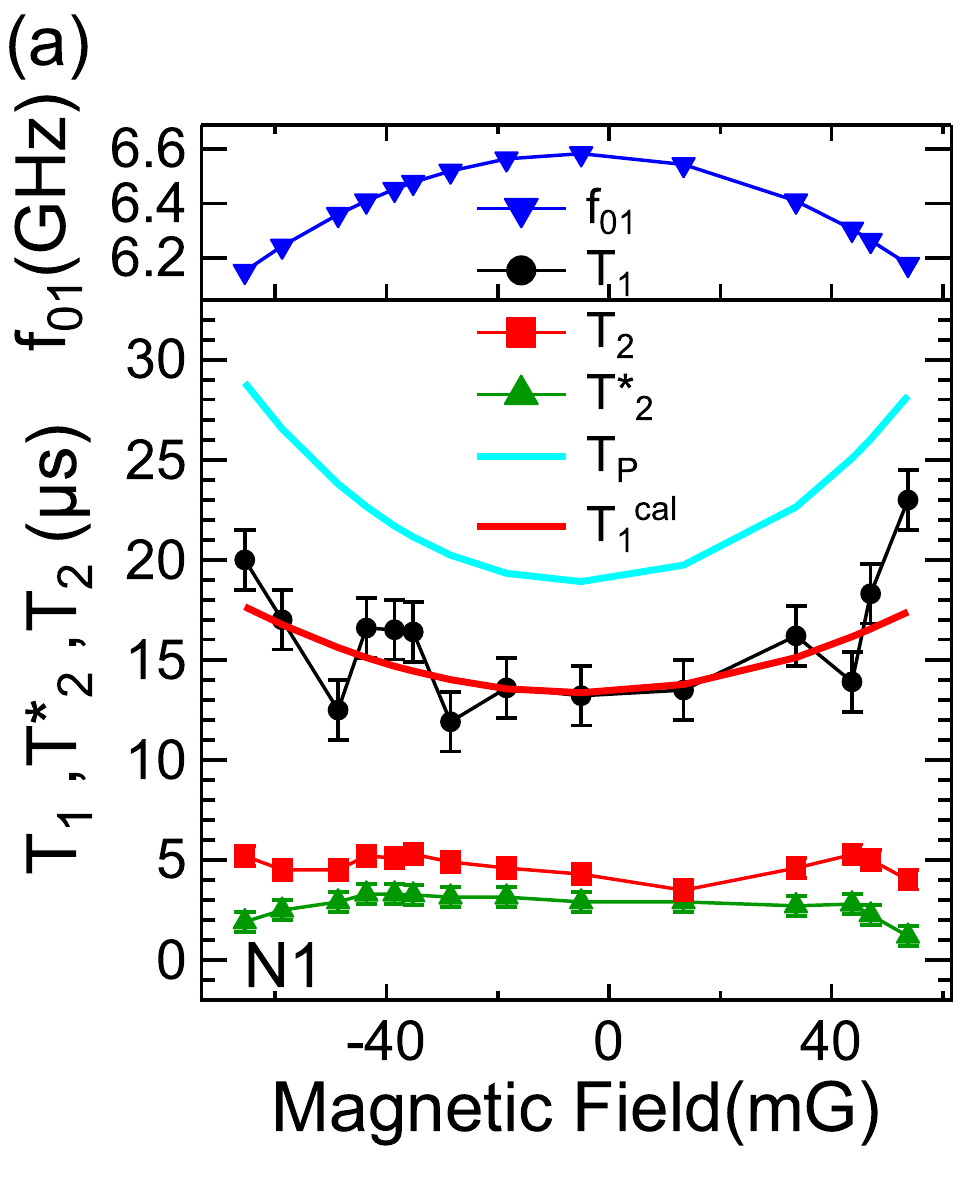}
\end{subfloat}
\begin{subfloat}{\label{fig:lowB_N7}}
\includegraphics[width=0.5\columnwidth]{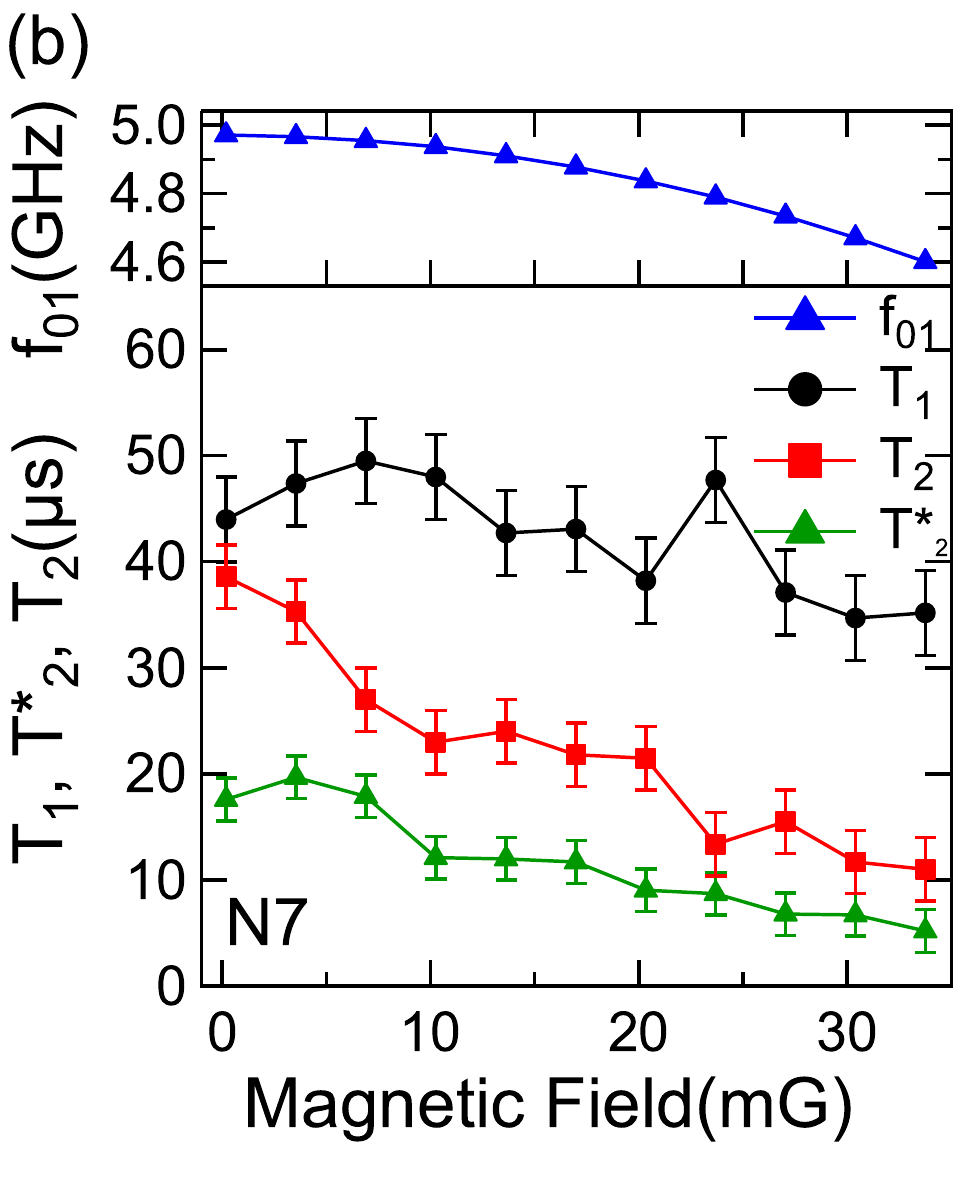}
\end{subfloat}
}
\caption{\label{fig:lowB}(Color online) (a) Magnetic field dependence of the qubit frequency ($f_{01}$), three measured time scales ($T_1$,$T^*_2$, and $T_2$) and two calculated time scales ($T_\text{P}$ and $T_1^\text{cal}$) at low magnetic field much smaller than the SQUID oscillation period for the sample N1. $T_\text{P}$ (Purcell time) was calculated by $T_\text{P}=1/\Gamma_\text{P}$---inverse of Purcell rate, and $T_1^\text{cal}$ by $1/T_1^\text{cal} =1/T_\text{NP}+1/T_\text{P}$ (see texts) (b) The qubit frequency and three measured time scales for N7.}
\end{figure}

Three time scales, $T_1$, $T^*_2$ and $T_2$ were extracted by fitting data with exponential decay function $\exp(-t/t_0)$ for $T_1$ and $T_2$, and sine-damped function $\exp(-t/t_0)\sin\left(2\pi f_{\text{\tiny R}}t+\varphi_0\right)$ for $T^*_2$. In Fig.~\hyperref[fig:lowB]{\ref{fig:lowB}}, we present $f_{01}$, $T_1$, $T^*_2$ and $T_2$ versus magnetic field measured in a vicinity of the sweet spot at $B$=0. The used weak magnetic field, up to $\sim$50 mG, is much weaker than the field needed to drive vortices into the electrodes. Thus here we discuss the vortex-free regime. 

We first examined the energy relaxation time, $T_1$, for both samples. They were measured separately in
the same cavity which has its loaded lowest order mode at $\sim$8.42 GHz with a loaded (measured) quality factor $Q_\text{L}=5000$. As Fig.~\hyperref[fig:lowB]{\ref{fig:lowB_N1}} and \hyperref[fig:lowB]{\ref{fig:lowB_N7}} show, at zero field the relaxation time $T_1$ was substantially larger for the sample N7 than for N1. Furthermore, when a small magnetic field was applied, the energy relaxation time for N1 increased, while hardly any change
was observed for N7. Both of these effects can be understood as consequences of the Purcell effect in
which the rate of spontaneous emission is increased when the cavity mode to which the qubit couples lies
close by in frequency. The excitation frequency of N7 (4.97 GHz) was further from the cavity frequency
than was the excitation frequency of N1 (6.583 GHz) and this almost completely determines the difference
in $T_1$. To see this we first compare the measured ratio $\Gamma_\text{N1}/\Gamma_\text{N7}$ and the calculated ratio of Purcell rates for the two devices. (Since the next higher cavity resonance is more than 11 GHz above the fundamental, we ignore its contribution to the Purcell relaxation rate.) For the Purcell relaxation rate we use $\Gamma_\text{P}=\kappa_1(g/\Delta)^2$, where $\kappa_1=\omega_c/Q_\text{L}$ is the cavity power decay rate, $g$ is the qubit-cavity coupling rate, and $\Delta$ is the qubit-cavity
detuning $\Delta=|\omega_{01}-\omega_c|=2\pi |f_{01}-f_\text{c}|$. The ratio of the Purcell rates depends only on the qubit-cavity frequency differences which are easy to measure. We find that the ratio of the qubits measured lifetimes for samples N1 and N7 is $T_\text{1,N1}/T_\text{1,N7}=13$ $\mu\text{s}/44$ $ \mu\text{s}=0.3$, while the ratio of the calculated Purcell times, $T_\text{P}=1/\Gamma_\text{P}$ is
$(\Delta_\text{N1}/\Delta_\text{N7})^2=0.29$. Since the measured and the estimated ratios are very close to each other, one can conclude that the relaxation is Purcell limited.

To confirm this conclusion we analyze the field dependence of the relaxation time. This is done using the qubit-cavity coupling, $g=130$ MHz (see Appendix~\hyperref[sec:device]{\ref{sec:device}}) and the formula for the Purcell rate $\Gamma_\text{P}=\kappa_1(g/\Delta)^2$. The increase in relaxation time for N1 with the applied B-field can be understood as a consequences of the Purcell effect. Indeed $\Gamma_\text{P}\sim1/\Delta^2$ decreases with $B$ because the qubit frequency decreases with increasing B-field thus causing the detuning $\Delta$ to increase. This makes a
measurable difference in $\Gamma_\text{P}$ for N1. For N1 we plot the relaxation time versus the B-field calculated as $1/T_1(B)=\Gamma_\text{P} (B)+\Gamma_\text{NP}$ [Fig.~\hyperref[fig:lowB_N1]{\ref{fig:lowB_N1}}]. Here $\Gamma_\text{NP}=2\pi\times 3.5$ kHz is a constant Non-Purcell rate estimated at zero magnetic field. A good agreement between the data and the fit is observed, confirming that the qubit is Purcell limited. Note that sample N7 does not show a noticeable dependence on the B-field because its detuning value is high and therefore the Purcell effect is relatively weak.

Finally we can estimate the internal relaxation rate of our devices assuming that the Purcell effect is eliminated by making the difference between the qubit frequency and the cavity difference sufficiently large. Such internal relaxation is represented by $\Gamma_\text{NP}$, as explained above. For N1 we obtain, at the sweet spot, $\Gamma_\text{P}=2\pi\times 8.5$ kHz and $\Gamma_\text{NP}=2\pi\times 3.5$ kHz, and, correspondingly $T_\text{NP}= 45$ $\mu\text{s}$. For N7 we estimate, again at the zero field sweet spot, $\Gamma_\text{P}=2\pi\times 2.4$ kHz and $\Gamma_\text{NP}=2\pi\times 1.2$ kHz, and, therefore $T_\text{NP}=132$ $ \mu\text{s}$. This analysis reveals that the relaxation time could be above 100 $\mu$s if it were not Purcell-limited, indicating that the coupling to energy absorbing defects in the circuit and qubit is low.

We now consider the spin echo coherence time $T_2$ at $B=0$ which measures the phase coherence attainable in a qubit process. $T_2$ is related to $T_1$ by the constitutive relation $1/T_2=1/(2T_1)+ 1/T_{\varphi}$, where $T_{\varphi}$ is the dephasing time due to random fluctuations of the phase evolution rate of the qubit wave function. N1 had a much shorter $T_2$ than N7. Specifically,
$T_\text{2,N1}=4.2$ $\mu\text{s}$ and $T_\text{2,N7}=39$ $ \mu\text{s}$, almost ten times as large. Using the measured values of $T_1$ we obtain
$T_{\varphi,\text{N1}}=3$ $\mu\text{s}$ and $T_{\varphi,\text{N7}}=70$ $\mu\text{s}$. We attribute the much longer dephasing time for N7 to the fact that the testing conditions were different. For measurements of N7, base-temperature copper powder filters were added to the input and output ports of the cavity. These are known to reduce stray-photon noise by providing attenuation at low temperature. Such photon noise may be responsible for the significantly lower dephasing time seen in the N1 measurements. The photon noise can induce dephasing because of a strong ac-Stark shift~\cite{Sears2012PRB}. Meanwhile, the $T_2$ of sample N7 became shorter as magnetic field was applied. Such behavior can be expected because an application of even a small $B$-field leads to a shift from the magnetic sweet spot. Thus the qubit becomes
more susceptible to dephasing caused by flux noise. This effect is less visible in the sample N1 because $T_2$ is already strongly suppressed by the stray-photon noise in this sample.

\subsection{Time domain measurement: high magnetic field}
Now we consider a different regime where sufficiently high magnetic fields creates vortices on the electrodes. We investigate the effect of Abrikosov vortices on the coherence times. Since the samples were zero-field-cooled there were no vortices in the electrodes to begin with. We managed to gradually increase the number of vortices by sweeping up the external perpendicular magnetic field. All measurements presented below have been made at the sweet spots, which occur periodically or approximately periodically with the magnetic field.

In contrary to a single transmon, the Meissner transmons have advantage to allow us to detect the entrance of a vortex (or multiple vortices) into the electrodes. In actual measurements, we ramped up the magnetic field until the next sweet spot was reached [see Fig.~\hyperref[fig:Bdep]{\ref{fig:TransmVsB}}] and then performed the next series of time-domain measurements. Upon the event of vortex entrance, we observed two signatures: hysteresis of the transmission plotted versus the magnetic field and a shift of the next sweet spot to higher magnetic field than would be expected if the pattern was exactly periodic. This happens because each vortex in the electrodes adds a phase gradient opposite to the one generated by the Meissner current [Fig.~\hyperref[fig:Bdep]{\ref{fig:TransmVsB}}]. 

\begin{figure}[b]
\makebox[\linewidth][c]{%
\begin{subfloat}{\label{fig:highB_N1}}
\includegraphics[width=0.5\columnwidth]{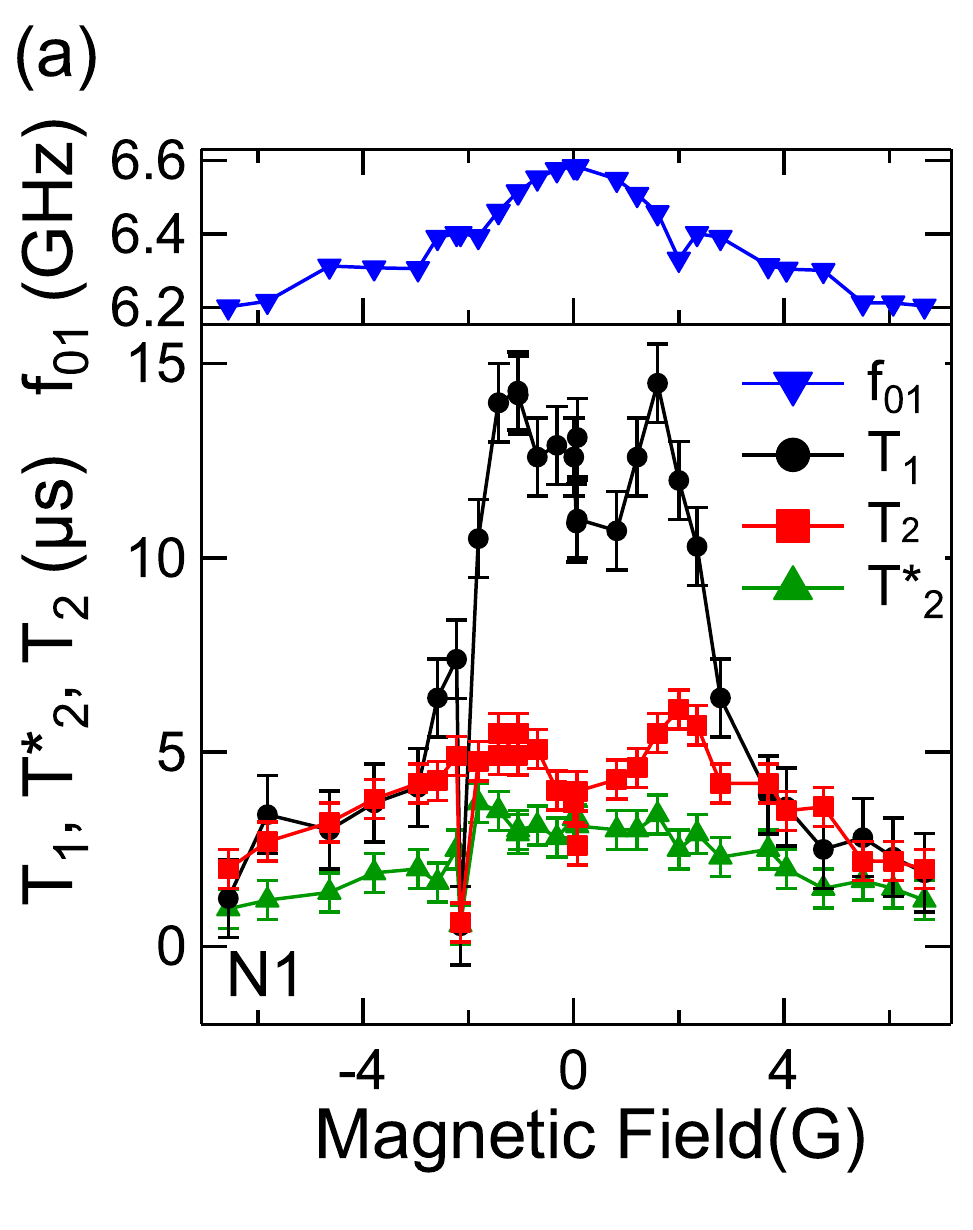}
\end{subfloat}
\begin{subfloat}{\label{fig:highB_N7}}
\includegraphics[width=0.5\columnwidth]{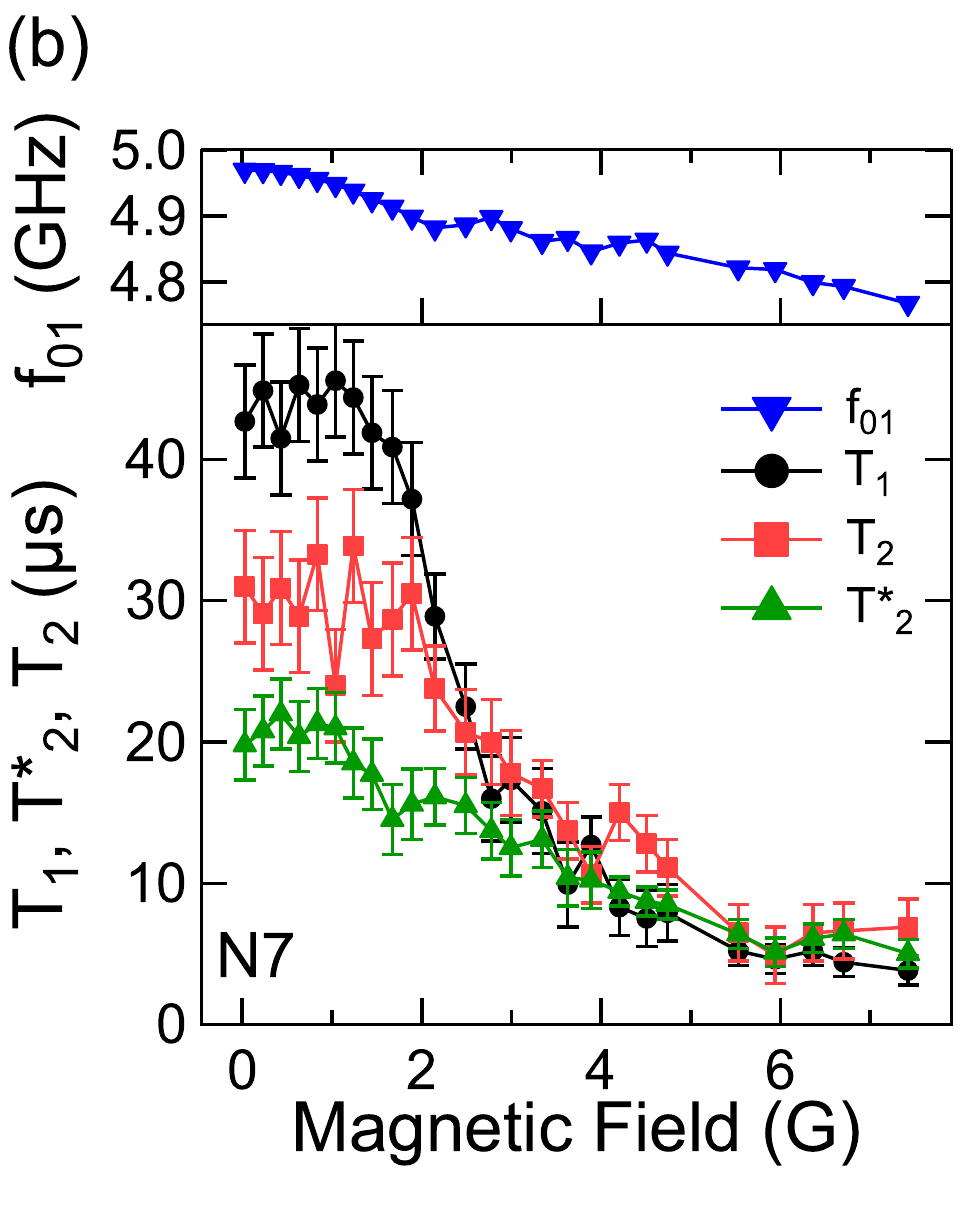}
\end{subfloat}
}
\caption{\label{fig:highB}(Color online) The qubit transition frequencies ($f_{01}$) and three times scales ($T_1$,$T^*_2$, and $T_2$) were measured at the sweet spots over the wide range of magnetic field for the N1 (a) and N7 (b). }
\end{figure}

In Fig.~\hyperref[fig:highB]{\ref{fig:highB_N1}}, we show the magnetic field dependence of three times scales and the qubit frequency for sample N1. It is observed that $T_1$ was enhanced from 10 $\mu$s to 14 $\mu$s as the magnetic field was increased. The trend was observed up to about 2 Gauss, while at higher fields the trend was reversed. In the case of sample N7 [Fig.~\hyperref[fig:highB]{\ref{fig:lowB_N7}}], all three time scales stayed almost constant up to about 2 Gauss and started to drop as the field was increasing further. This is explained by the fact that vortices begin to penetrate into the electrodes at the first critical field $B_\text{c1}\approx$2 G. They provide a radiation-free dissipation source and thus suppress the relaxation time significantly [see Fig.~\hyperref[fig:highB]{\ref{fig:highB}}]. Meanwhile, the coherence time $T_2$ (and $T^*_2$) also decreased, mainly due to the reduction of $T_1$. Since the measurements were carried out at the sweet spots, the dephasing caused by non-zero dispersion of $f_{01}(B)$ was negligible. 

The entrance of vortices is confirmed by a comparison to a theoretical model. In Ref.~\onlinecite{Marksimov1997} the first critical field is approximated as $B_\text{c1}$=$\Phi_0/\left[2\pi\xi\sqrt{2\lambda_{\perp} X}\right]$, where $\lambda_{\perp}$=$2\lambda(0)^2/d$ represents the penetration depth of a thin film in perpendicular field. Using the relations, $\lambda(0)$=$ \lambda_\text{L}\sqrt{1+\xi_0/l}$ and $\xi$=$\sqrt{\xi_0 l}$, we estimate $B_\text{c1}$=5.9 G which is similar to the measured value, 2 G. Here $\xi$ is the coherence length, $\xi_0$ is the clean limit coherence length, $\lambda (0)$ is the bulk penetration depth, $ \lambda_\text{L}$ is the clean-limit penetration depth for Al. The parameter values are: $\xi_0=1600$ nm, $l=16.7$ nm, $ \lambda_\text{L}=16 $ nm, $\xi$=163 nm, $\lambda (0)=158$ nm, $\lambda_{\perp}=552$ nm, $X=10\,\mu$m. The electronic mean free path $l$ is calculated from the measured resistivity of the Al films forming the electrodes, $\rho_\text{n}=2.4\times 10^{-8}$, according to Ref.~\cite{Romijn1982PRB}, using $l\rho_\text{n}=4\times10^{-16}\,\Omega$m$^2$.

\begin{figure}[t,b]
\makebox[\linewidth][c]{%
\begin{subfloat}{\label{fig:T1rate_N1}}
\includegraphics[width=0.5\columnwidth]{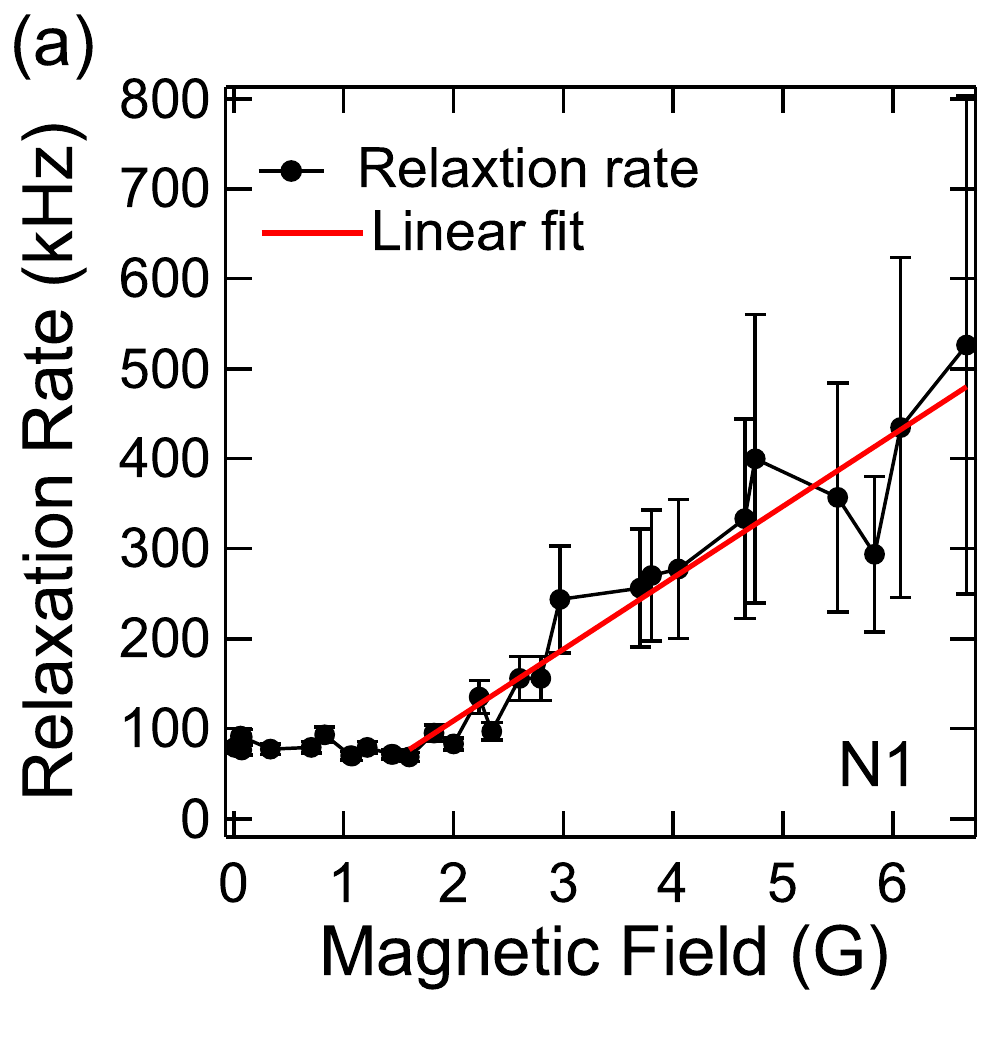}
\end{subfloat}
\begin{subfloat}{\label{fig:T1rate_N7}}
\includegraphics[width=0.5\columnwidth]{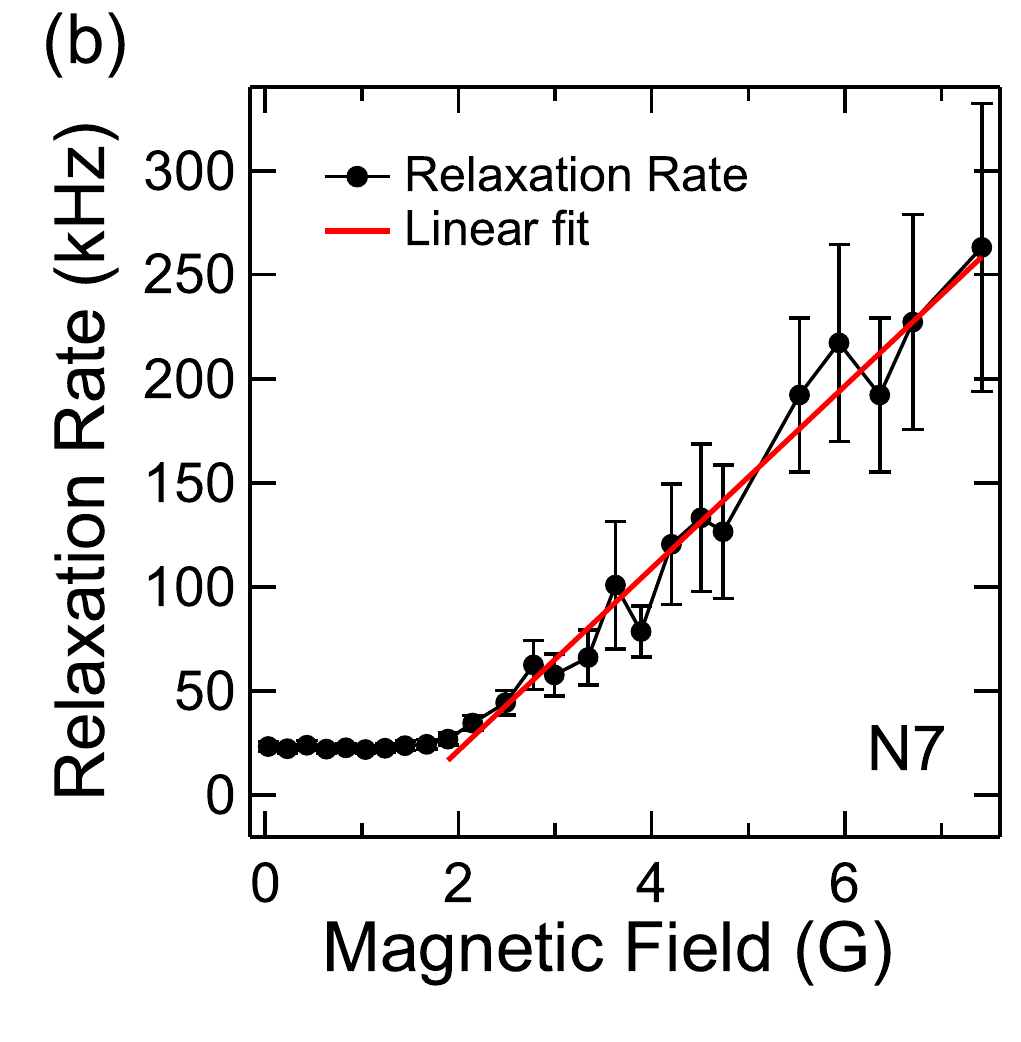}
\end{subfloat}
}
\caption{\label{fig:T1rate}(Color online) The relaxation rates $\Gamma$=$1/T_1$ versus magnetic field were plotted for both N1 (a) and N7 (b). All measurements have been done at the sweet spots. The linear fits (red solid lines) were shown over the magnetic field range where the vortices were present in the electrodes. }
\end{figure}  

Our goal now is to achieve a quantitative characterization of the non-radiative relaxation process caused by vortices. For this, we plot the relaxation rate $\Gamma=1/T_1$ versus magnetic field, $B$, in Fig.~\hyperref[fig:T1rate]{\ref{fig:T1rate}}. The relaxation rate remains approximately constant at $B<B_\text{c1}$ and increases approximately linearly at $B>B_\text{c1}$. The observed increase of the relaxation rate per Gauss ($d\Gamma/dB$) was 78.5 kHz/G for sample N1 and  43.7 kHz/G for sample N7. In what follows we suggest a model of non-radiative decay, which can explain these values.

We suggest that the energy relaxation of the qubits is mainly due to the energy dissipation originating from the vortex viscous motion. The motion of vortices is initiated be the Lorentz force, which, in turn, is due to the currents generated by the qubit itself. The motion of vortices is overdamped due to the viscous drag force, which, per unit vortex length, is $\textbf{f}_v=-\zeta\textbf{v}$, where $\zeta$ is a viscous drag coefficient for a vortex of unit length, and $\textbf{v}$ is the vortex velocity~\cite{Tinkhambook}. This process is a non-radiative relaxation in which the qubit energy is dissipated as heat. We estimate this relaxation rate semi-classically, using the Bardeen-Stephen model~\cite{Tinkhambook}. According to their model, the viscosity $\zeta$ per unit length is $\zeta=\Phi_0B_\text{c2}/\rho_\text{n}$, where $B_{c2}=\Phi_0/(2\pi\xi^2)$, and $\xi=\sqrt{\xi_0 l}$. Here $B_{c2}$ is the second critical field of the thin-film Al electrodes and $\rho_\text{n}$ is their normal-state resistivity.

Let us estimate the energy relaxation rate, $\Gamma_\text{v}$, caused by one vortex via viscous damping. The rate of the energy dissipation---dissipated power---is $P=-(\textbf{f}_\text{v}\cdot\textbf{v}) d=f_\text{v}^2d / \zeta$, where $d$ is the thickness of film. Thus the energy relaxation rate of a transmon by a vortex can be evaluated as $\Gamma_\text{v}$=$P/(\hbar\omega_{01})$=$\left(f_\text{v}^2/\zeta\right)d/\hbar\omega_{01}$, where $\hbar\omega_{01}$ is the energy stored in the first excited state of the qubit. The vortex is driven by the Lorentz force $\textbf{f}_L$=$\textbf{J}\times\Phi_0$ (this is the force per unit length), where $\textbf{J}$ is a supercurrent density. The supercurrent density is proportional to the total current, $J=I/Xd$, i.e., the current density magnitude is approximated by the total current $I$ divided by the cross section area of the electrode. A naive first guess might be that vortices in the electrodes should not move at all since the expectation value of the current generated by the qubit is zero, $\langle J\rangle =0$, both in the ground and in the exited state of the qubit. Yet we will see soon that the dissipation is proportional to  $\langle J^2\rangle$, which is greater than zero.

The next step is to set $f_\text{v}=f_\text{L}$ based on a reasonable assumption that the effective mass of the vortex and the pinning force are negligible. Consequently, we obtained the energy relaxation rate per vortex, 
\begin{equation}\label{eq:dissipationrate}
\Gamma_\text{v}=\frac{J^2\Phi^2_0 d}{\zeta \hbar\omega_{01}},
\end{equation}

Of course for the quantum states of the qubit the current and the current density are quantum variables which do not have definite values but should be viewed as quantum mechanical operators. The probability amplitudes of these quantities are defined by the wave function of the qubit. Thus under $J^2$ we understand the mean square of the current density, $J^2=\left<1|\hat{I}^2|1\right>/(Xd)^2$, where the averaging is done for the excited quantum state of the qubit. Here $\hat{I}$ is the operator of the current in the qubit and $X$ and $d$ represent the width and the thickness of the electrodes. 

The model outlined above leads to the following estimates for the relaxation rate induced on the qubit by a single vortex: $\Gamma_\text{v}=89$ kHz/vortex and $\Gamma_\text{v}=48$ kHz/vortex, for samples N1 and N7 correspondingly. The following set of sample-specific parameters has been used for sample N1, $\sqrt{\left<I^2\right>}=29.5$ nA, $J=32.8$ kA/m$^2$, $\omega_{01}/2\pi=6.583$ GHz, $\zeta=1.1\times 10^{-9}$ N$\cdot$s/m$^2$, $B_{c2}=12.2$ mT, $\rho_\text{n}=2.4\times 10^{-8}$ $\Omega\cdot$m, $\xi =163$ nm, $\xi_0=1600$ nm, $X=10$ $\mu$m, $d=90$ nm, and $l =16.7$ nm. For sample N7, all the parameters are the same, with the exception of  $\sqrt{\left<I^2\right>}=18.9$ nA, $J=21.0$ kA/m$^2$, and $\omega_{01}/2\pi=4.972$ GHz.

\begin{figure}[t,b]
 \includegraphics[width=0.8\columnwidth]{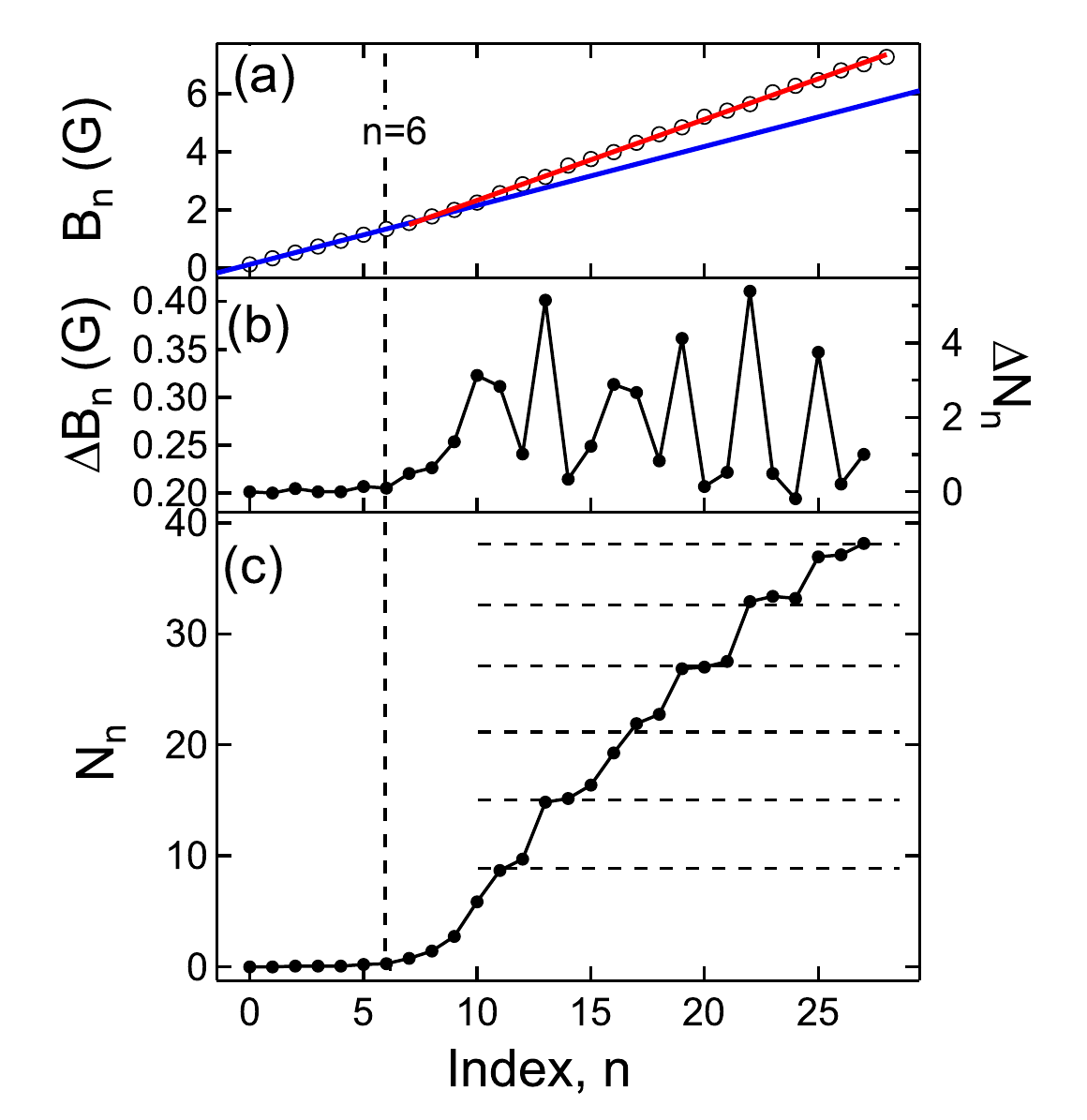}
\begin{subfloat}{\label{fig:vortexcount_a}}
\end{subfloat}
\begin{subfloat}{\label{fig:vortexcount_b}}
\end{subfloat}
\begin{subfloat}{\label{fig:vortexcount_c}}
\end{subfloat}
\caption{(Color online)(a) The magnetic fields $B_n$ at sweet spots (black open circle) are depicted as a function of $n$---index of the sweet spots, for sample N7. The blue  and red solid lines represent a linear fit to the data for $n\leq6$ and $n\geq7$, respectively. The blue fitted line is extended for $n\geq7$ to show the expected $B_n$ when no vortex penetration is assumed. (b) The difference $\Delta B_n$=$B_n-B_{n-1}$ is shown on the left axis, while the number of vorticies $\Delta N_n$ (defined in the text) provides the scale for the right axis. (c) The total number of vorticies is calculated by summation of $\Delta N_n$. The dashed lines are periodically spaced. They provide a guide-to-the-eye to emphasize the stepwise characteristics of the increasing number of vortices in the electrodes.}\label{fig:vortexcount}
\end{figure} 

To compare the relaxation rate $\Gamma_\text{v}$, computed per a single vortex (see above), with the experimental relaxation rates, $d\Gamma/dB$, measured ``per Gauss'', we need to estimate the average rate of the vortex entrance, $dN/dB$. Then one can use a formula $d\Gamma/dB=\Gamma_\text{v}(dN/dB)$, which assumes that the relaxation rates of all vortices simply add up. 

A detailed analysis of various possibilities to estimate the vortex entrance rate are given in Appendix~\hyperref[sec:vortex_count]{\ref{sec:vortex_count}}. Here we briefly outline the most intuitive estimate. First, we define $B_n$ as the sequence of magnetic fields corresponding to the sequence of the sweet spots, indexed by the integer $n$=0, 1,$\ldots$, 27. Here 27 is the total number of the observed sweet spots [see Fig.~\hyperref[fig:vortexcount]{\ref{fig:vortexcount_a}}]. 

At low fields $B_n$ (black open circle) increase linearly with $n$, due to exact periodicity of the HV-oscillation in the vortex-free regime. The linear fit [blue line in Fig.~\hyperref[fig:vortexcount]{\ref{fig:vortexcount_a}}] gives the value of the period, $\Delta B$ =0.2 G. For $n>7$ the period becomes larger because vortices begin to penetrate. The new slope, and, correspondingly, the new period is $\Delta B+\Delta B_\text{v}=0.278$ G, for sample N7. This best-fit value is obtained from the linear fit represented by the red solid line in Fig.~\hyperref[fig:vortexcount]{\ref{fig:vortexcount_a}}. The enlargement of the period happens because vortices compensate, to some extent, the strength of the Meissner current. 

Now we calculate the difference between the consecutive sweet spot fields $\Delta B_n$=$B_n-B_{n-1}$. The result is plotted in Fig.~\hyperref[fig:vortexcount]{\ref{fig:vortexcount_b}}. The result was then converted into some effective change of the vortex number, $\Delta N \sim (B_n-B_{n-1})-\Delta B$, where $\Delta B$ is the distance between the sweet spots in the vortex-free low-field regime. Thus defined $\Delta N$ should be considered as a function proportional to the number of vortices coupled to the qubit. But since the conversion factor is not well known, $\Delta N$ should not be considered as equal to the number of relevant vortices. Finally, we integrate $\Delta N_{n}$ with respect to $n$. The result is shown in Fig.~\hyperref[fig:vortexcount]{\ref{fig:vortexcount_c}}. This integrated function exhibits clear steps, which we interpret as vortices entering the sensitivity area of the qubit. The steps are made more noticeable by placing the horizontal dashed lines. The spacing between the lines is made constant and they serve as guide to the eye.  The total number of effectively coupled vortices equals the number of steps, i.e. equals 6. These 6 vortices have entered over the interval of 5.6 Gauss. Thus the effective entrance rate is estimated as $dN/dB\approx$ 1.07 vortex/G. 

Remember that the experimental relaxation rates per Gauss $d\Gamma/dB$, obtained from Fig.~\hyperref[fig:T1rate]{\ref{fig:T1rate}}, are 78.5 and 43.7 kHz/G for samples N1 and N7. Now these values need to be divided by 1.07 vortex/G, which is the rate of the vortex entrance. Thus we conclude that the experimental relaxation rates are $\Gamma_\text{v}=73$ kHz/vortex and $\Gamma_\text{v}=41$ kHz/vortex, for samples N1 and N7, calculated using $\Gamma_\text{v}=(d\Gamma/dB)/(dN/dB)$. These values are in excellent agreement with the theoretical estimates 89 kHz and 48 kHz/vortex. 

To understand the significance of the obtained results for the macroscopic quantum physics, it is instructive to compare the relaxation rates generated by individual vortices to the theory of localization caused by dissipative environment. According to the Caldeira and Leggett (CL) theory~\cite{Caldeira1981PRL}, the particle wave function should be exponentially localized with the localization length scale estimated as $x_\text{CL}^2\simeq h/\eta$, where $\eta$ is the viscous drag coefficient of the macroscopic quantum particle coupled to the environment. This theory was later generalized to the case of periodic potentials in Refs.~\cite{SchmidPRL1983,BulgadaevJTEPL1984}. The conclusion of these papers was that if the period of the potential is larger than $x_\text{CL}$ then the particle becomes localized in one of the wells in the limit of zero temperature. On the other hand, if the period is smaller than $x_\text{CL}$ then the particle can tunnel from one minimum to the next one even at zero temperature. It is important that the scale of the environmental localization, $x_\text{CL}$, is independent of the amplitude of the modulation of the potential energy. Thus it can provide a useful estimate even if the potential is approximately flat, which is the case for the Abrikosov vortex in the Al film in our devices. The viscous drag coefficient for a single vortex is $\eta=d\zeta=9.9\times 10^{-17}$ N$\cdot$s/m, where $d=90$ nm is the vortex length, which is set by the film thickness $d$. Then the Caldeira-Leggett (CL) localization scale is $x_\text{CL}=\sqrt{h/\eta}=2.6\times 10^{-9}$ m = 2.6 nm. This is the scale up to which the wave function of a vortex can spread. If the spread is larger than $x_\text{CL}$ then the coupling to the environment causes the wave function collapse. 

The CL localization scale should be compared to the estimated smearing of the wave function of the center-of-mass of the vortex, generated by fluctuations of the supercurrent and the corresponding fluctuations of the Lorentz force. The smearing of the wave function can be estimated as follows. The root-mean-square (rms) value of the Lorentz force is $F_\text{L}=d\cdot f_\text{L}=6.1\times 10^{-18}$ N. (Here we consider the example of sample N1. All estimates for sample N7 are very similar.). Then, assuming viscous motion, the rms velocity is $v=F_\text{L}/\eta=6.2\times10^{-2}$ m/s. Therefore, the rms quantum fluctuations of the vortex center position, $x_\text{v}$, can be estimated as $x_\text{v}=v/\omega_{01}$. This relation would be exact if the motion of the vortex would be described by a classical trajectory, in response to a harmonic drive, such that the deviations from its point of equilibrium would be proportional to $\sin(\omega_{01}t)$. In the case considered the vortex is not described by a classical trajectory, since it should act as a quantum particle at time scales shorter than the relaxation time of the qubit and also because the force is generated by a quantum superposition of currents with opposite polarities. But we assume that the relationship between the quantum fluctuation of the position and the velocity is approximately the same as in the case of a classical harmonic motion. Such assumption is motivated by the natural expectation that the vortex should behave as a damped quasi-classical particle. Thus we can now evaluate the rms smearing, $x_\text{v}$, of the wave function of the vortex center induced by the quantum fluctuations of the Lorentz force. The result is $x_\text{v}=(F_\text{L}/\eta)/\omega_{01}=1.5\times 10^{-12}$ m = 1.5 pm. Here we have used the qubit frequency $f_{01}=\omega_{01}/2\pi=6.58$ GHz for sample N1. The conclusion is that the quantum uncertainty of the vortex position is much smaller than the CL localization scale, $x_\text{v} \ll x_\text{CL}$, namely $x_\text{CL}/x_\text{v}=1700$. Such uncertainty is achieved within one period, $T_{01}=2\pi/\omega_{01}$.

The CL scale provide the maximum value of the rms smearing of the wave function. When such level of smearing is achieved, the wave function collapses and the qubit experiences a decoherence event. The number of complete phase revolutions of the qubit, $N_\text{dch}$, which is needed to achieve the CL scale, at which the probability of decoherence becomes of the order of unity, can be estimated assuming that the wave function of the vortex spreads similar to a diffusive random walk. Then $N_\text{dch}=x_\text{CL}^2/x^2=3.0\times 10^6$. Thus the corresponding decoherence time can be estimated as $t_\text{dch}\sim N_\text{dch}T_{01}$. This heuristic argument can be made more precise using Eq.~\eqref{eq:dissipationrate}. From that equation, using the relations listed above, one obtains $t_\text{dch}=(1/2\pi)^2 N_\text{dch}T_{01}=12\,\mu$s. Finally, the estimate for the relaxation rate, added to the system due to one vortex coupled to the qubit, is $\Gamma_\text{v}=1/t_\text{dch}=83$ kHz. This analysis leads to a conclusion that the relaxation rate induced by one vortex can be linked to the ratio of the smearing of its wave function and the Caldeira-Leggett dissipative localization scale. The above estimate shows that the wave function collapse occurs typically at the time when the wave function smearing reaches the Caldeira-Leggett localization scale.    

\section{Summary}\label{sec:summary}
We demonstrate the operation of Meissner qubit, which is a transmon qubit strongly coupled to the Meissner current in the electrodes. The periodicity in magnetic field was set mainly by the width of the electrodes rather than the SQUID loop area. Both the frequency and the time domain measurements were performed using the circuit-QED architecture. The three time scales of $T_1$, $T_2^*$ and $T_2$ were measured as functions of the applied magnetic field. The increase of the relaxation rates was attributed to the radiation-free dissipation associated with the viscous motion of Abrikosov vortices pushed by fluctuations of the Lorentz force. Each vortex can exist in a quantum superposition of different position states for $\sim$10-100 $\mu$s, but eventually causes the wave function of the coupled qubit to collapse. The collapse happens when the smearing of the vortex center becomes of the order of the Caldeira-Leggett dissipative localization scale. The presented Meissner qubit provides an effective and controlled coupling of the qubit to Abrikosov vortices. Such coupling provides a new tool to study vortices, which can eventually be applied to vortices harboring Majorana states. 

\begin{acknowledgments}
This work was supported by the National Science Foundation under the Grant No. ECCS-1408558 (A.B.) and ECCS-1407875 (A.L.).
\end{acknowledgments}

\appendix

\section{Device and Measurement} \label{sec:device}
The Meissner qubits were fabricated on a c-plane sapphire using a novel modification of the double-angle evaporation technique, now achieved in ultra-high vacuum in a molecular beam epitaxy (MBE) growth system. The device design is shown in Fig.~\hyperref[fig:setup]{\ref{fig:setup_d}}. The patterns were defined by electron beam lithography on the bi-layer of MMA EL-13 and ZEP 520 A7 in eLine Raith system, and after development the exposed surface was cleaned to remove MMA residue by both dry and wet etching---oxygen plasma by RIE (reactive-ion etching) and BOE (buffered oxide etch). The first and second layers of aluminum films, each with 45 nm thickness were deposited with background  base pressure of $10^{-11}$ Torr. The oxide layer was formed by an exposure to Ar/O$_2$ mixture (10$\%$ O$_2$) under the proper conditions of pressure and time calibrated for critical current density of the JJs. Each large rectangle [marked A1 and A2 in Fig.~\hyperref[fig:setup]{\ref{fig:setup_b}}] acts as a radio-frequency (RF) antenna and has dimension of $250\times500\,\mu\text{m}^2$. The spacing between the nearest edges of the antennas is 25 $\mu$m [Fig.~\hyperref[fig:setup]{\ref{fig:setup_b}}]. The antennas are bridged by two Al thin-film rectangles [called ``electrodes'' and marked E1 and E2 in Fig.~\hyperref[fig:setup]{\ref{fig:setup_c}}] and two JJs, connecting the electrodes and forming a SQUID-like loop [Figs.~\hyperref[fig:setup]{\ref{fig:setup_c},\ref{fig:setup_d}}]. 

We will discuss two representative devices denoted by N1 and N7. The qubit transition frequency $f_{01}$ of N1 at zero magnetic field was 6.583 GHz, the Josephson energy of both junctions taken together was $E_\text{J}^\text{max}$=19.4	GHz, the corresponding net critical current was 41 nA, and the Coulomb charging energy---mostly associated with the electric capacitance between the antennas---was $E_\text{C}$=0.307 GHz. Thus the ratio of the two energy scales was $E_\text{J}^\text{max}/E_\text{C}$=63.2 for device N1. For the qubit N7 the same type of parameters were $f_{01}$=4.970 GHz, $E_\text{J}^\text{max}$=11.1 GHz, the corresponding critical current at zero field was 23.4 nA, $E_\text{C}$=0.318 GHz, and $E_\text{J}^\text{max}/E_\text{C}$=34.9. 

Each device was mounted in a 3D rectangular copper cavity with the dimension $28\times 22\times 5\,\text{mm}^3$ [Fig.~\hyperref[fig:setup]{\ref{fig:setup_a}}]. The qubits were coupled to electromagnetic cavity modes and in turn the cavity was used for qubit readout. The samples were positioned in the center of the cavity where the dipole coupling of the fundamental mode to the qubit is maximized, i.e., at the electric field anti-node. The TE$_{011}$ mode of the empty cavity occurred at 8.679 GHz with the internal (unloaded) quality factor $Q_\text{i}$=8000, external quality factor $Q_\text{e}$=14000, and loaded ("experimental") quality factor $Q_\text{L}$=5000~\cite{Qfactor}. An asymmetric coupling (larger at output) was used to maximize the signal to noise ratio~\cite{noiseratio}. The bare cavity mode $f_\text{c}$~\cite{barefreq} at TE$_{011}$ for sample N1(N7) was 8.403 GHz (8.435 GHz), and the coupling strength $g/2\pi$ was approximately 130 MHz~\cite{couplingstrength} for both samples. 

The cavity was mounted in a $^3$He/$^4$He dilution refrigerator (S.H.E. Corp.) with the base temperature of 45 mK. The cavity was enclosed in a cylindrical aluminum Faraday cage, whose inner walls were coated with black infrared-absorbing material~\cite{blackmagic}. The Al cage is intended to prevent external stray photons from reaching the sample. The Al cage helps also to protect the sample from the influence of external stray magnetic fields, thanks to the Meissner effect. A cylindrical cryogenic $\mu$-metal (Amuneal) cylinder was placed concentrically around the Al cage for additional magnetic shielding. The magnetic field was applied perpendicular to the substrate from an external home-made superconducting solenoid attached at the bottom of the copper cavity, i.e. inside the Al cage and the $\mu$-metal shield.

For microwave transmission measurement, the input and output transmission lines were connected in series with a chain of cryogenic microwave components, including attenuators, isolators (PAMTEQ), a commercial low-noise HEMT (high-electron-mobility transistor) amplifier (Low Noise Factory, LNF-LNC6-20A) and low-pass filters. For noise filtering, commercial low-pass filters (K$\&$L Microwave, 6L250-12000) and home-made stainless steel powder filters (3 dB at cutoff $\approx 8$ GHz) were inserted in both input and output of the copper cavity.

The measurements were performed using the circuit QED technique in high-power regime~\cite{Paik2011PRL, Reed2010PRL}. For the spectroscopy and qubit readout, RF square pulses were created, added together and fed into the cavity. The pulses were shaped by mixing a continuous microwave tone (Agilent E8267C or HP 8341B) and a square voltage pulse from an arbitrary waveform generator (Tektronix AWG520) using pairs of rf mixers (Markis, M8-0420). The transmitted readout signal was down-converted to 25 MHz intermediate frequency (IF) signal by heterodyne demodulation, and the IF signal was acquired to read the amplitude by a high-speed digitizer (Agilent, U1082A-001). For qubit state readout, we adjusted the power of readout pulse (a few $\mu$s long) to maximize the contrast in transmitted microwave amplitude for the ground and first excited states.

\section{Calculation of $\kappa$: Magnetic field focusing effect}\label{sec:kappa_cal}
 \begin{figure}[t]
\centering
 \includegraphics[width=0.6\columnwidth]{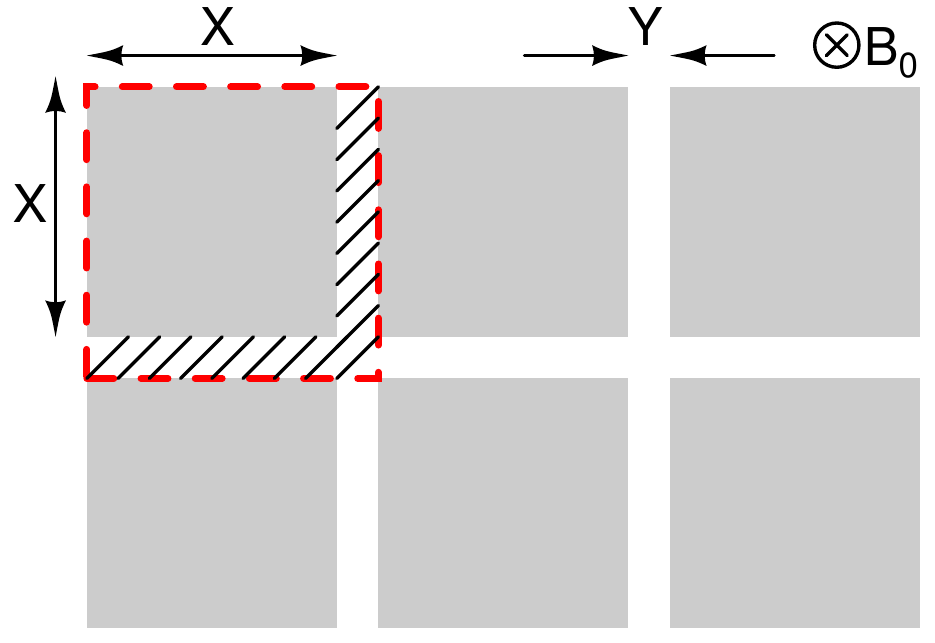}
\caption{\label{fig:fieldfocus} (Color online) An array of square superconducting aluminum films. A magnetic field is applied perpendicular to the surface of the films. The enclosed area by dashed line indicates one unit cell.}
\end{figure}
In this appendix, we will show how to calculate $\kappa$. Consider a two-dimensional array of square superconducting films with magnetic field $B_0$ applied perpendicular to the in-plane of the array as shown in Fig.~\hyperref[fig:fieldfocus]{\ref{fig:fieldfocus}}. The squares represent the electrodes in actual Meissner qubits. We consider a unit cell enclosed by a red dashed line to calculate $\kappa$. When the magnetic field $B_0$ is applied over the unit cell, the magnetic field inside the films is expelled by Meissner effect, and thus the magnetic field ($B_1$) in the hatched area is enhanced by a factor of $\kappa = B_1/B_0$. $\kappa$ is calculated in the following way. We denote $A_0$ to the area of a unit cell and $A_1$ the hatched area. The magnetic flux in one unit cell is $\Phi=B_0 A_0 = B_1 A_1$, so
\begin{equation}\label{eq:kappa}
\kappa = \frac{B_1}{B_0}=\frac{A_0}{A_1}=\frac{(X+Y)^2}{(X+Y)^2-X^2}=\frac{(X+Y)^2}{Y(2X+Y)} 
\end{equation}
We note that $\kappa > 1$, i.e., field focusing effect.

\section{Analysis of the average number of vortices entering the electrodes per one Gauss of the external field, $dN/dB$.}\label{sec:vortex_count}

It is not possible to determine $dN/dB$ exactly. Therefore we outline three different approaches below. The first two methods, which are very similar, will provide the higher bound for $dN/dB$, while the third one will give the lower bound for $dN/dB$. 

First, we define $B_n$ as the sequence of magnetic field values corresponding to consecutive sweet spots, indexed by the integer $n$=0, 1,$\ldots$, 27. Here 27 is the maximum number of the sweet spots measured on sample N7. The sweet spots correspond to the minima of the HV-signal shown in Fig.~\hyperref[fig:TransmVsB]{\ref{fig:TransmVsB}}. 

At low fields [for $n\leq 6$ in Fig.~\hyperref[fig:vortexcount]{\ref{fig:vortexcount_a}}], $B_n$ (black open circle) increase linearly with $n$, as is expected for the situation in which the sweet spots occur periodically with magnetic field. Such exact periodicity is observed only in the low field regime, when there are no vortices in the electrodes, i.e., $B_n<B_\text{c1}$ for $n\leq 6$. Since the oscillation is perfectly periodic in this vortex-free regime, the positions of the sweet spots of the qubit can be approximated as $B_n=\Delta B\cdot n+B_0$, where $\Delta B$ is the unperturbed period of the HV-oscillation and $B_0$ is the position of the zero's sweet spot. The linear fit [blue line in Fig.~\hyperref[fig:vortexcount]{\ref{fig:vortexcount_a}}] in the low-filed regime gives the value of the period, $\Delta B$ =0.2 G for $n\leq 6$. Note that in such representation the best linear fit provides the averaged period, $\Delta B =dB_n/dn$. 

In what follows we discuss the regime occurring above the critical field, with vortices entering the electrodes as the magnetic field is swept up. Since the slope of the $B_n$ versus $n$ dependence changes significantly at $n=7$, therefore $B_7\approx B_\text{c1}$. For $n>7$ the period is larger compared to the unperturbed case with no vortices. This is because the current of each vortex is opposite to the Meissner screening current. Thus the total phase bias imposed on the SQUID loop by the electrodes~\cite{Hopkins2005science} increases slower with the magnetic field if the number of vortices in the electrodes increases with magnetic field. Thus the sweet spots tend to occur at higher field values. The new dependence of the position of the sweet spots versus their consecutive number is still approximately linear [see Fig.~\hyperref[fig:vortexcount]{\ref{fig:vortexcount_a}}], but the slope is larger. The formula for the sweet spot sequence becomes $B_n\approx(\Delta B+\Delta B_\text{v})n+B_0^{'}$, where $\Delta B_\text{v}$ is the value by which the average interval between the sweet spots is increased, due to the continuous increase of the number of vortices $N$ in the electrodes. The new slope, and, correspondingly, the new period is $\Delta B+\Delta B_\text{v}=0.278$ G, in the present example of sample N7. This best-fit value is obtained from the linear fit represented by the red solid line in Fig.~\hyperref[fig:vortexcount]{\ref{fig:vortexcount_a}}. 

The enlargement of the period $\Delta B_\text{v}=0.078$ G is attributed to the additional phase bias~\cite{Golod2010PRL} induced on the SQUID loop by the vortices entering in the electrodes as the field is swept from one sweet spot to the next one. This change in the period can now be used to estimate the number of vortices entering the electrodes within one period of the HV-oscillation. Let us $\Delta\varphi_\text{v}$ be the average phase difference imposed by a vortex on the SQUID loop and $\Delta N$ be the average number of vortices entering the electrodes during each period. Then the additional phase difference $\Delta\varphi_\text{vv}$ accumulated during each period due to all newly entered vortices is $\Delta\varphi_\text{vv}=\Delta\varphi_\text{v}\cdot \Delta N$. Since the vortex current opposes the Meissner current, the phase difference $\Delta\varphi_\text{vv}$ is opposite in sign to the phase $\Delta\varphi_\text{M}$ imposed by the Meissner current. Therefore the total phase bias generated within one period can be written as $\Delta\varphi_\text{t}=\Delta\varphi_\text{M} - \Delta\varphi_\text{v}$. As in any SQUID-based device, one period of oscillation corresponds to the total phase change by $2\pi$. Thus one has to require $\Delta\varphi_\text{t}=2\pi$ and so $\Delta\varphi_\text{M} = 2\pi + \Delta\varphi_\text{v} \Delta N$. 

The value of $\Delta\varphi_\text{M}$ is determined by the phase-bias function $2\delta(B)$, which is the function defining how much phase bias is produced by the applied magnetic field B [see Eq.~\eqref{eq:phaseconstraint}], taking into account only the Meissner-current-generated phase gradients in the electrodes~\cite{Hopkins2005science,Pekker2005PRB}. Since in the vortex-free regime the period equals $\Delta B$ and the phase bias should change by $2\pi$ to complete one period, and also because $\delta B$ is a linear function of $B$, we can write 
\begin{equation}\label{deltaB}2\delta(B)=2\pi(B/\Delta B) 
\end{equation}
for the vortex-free regime. As vortices begin to enter the electrodes, the period increases, on average, to $\Delta B+\Delta B_\text{v}$, as was discussed above. Again, the total phase generated within one period should be $2\pi$. Thus, when vortices are entering we can write $2\delta(\Delta B+\Delta B_\text{v})= 2\pi + \Delta\varphi_\text{v} \Delta N$. The last term is needed since at the end of one period the phase generated by the Meissner current has to be larger than $2\pi$ by as much as $\Delta\varphi_\text{v} \Delta N$, to compensate the opposite phase bias, $\Delta\varphi_\text{v} \Delta N$, generated by the newly entered vortices. Remember that the function $\delta (B)$ is defined by Eq.~\eqref{deltaB}. Therefore $2\pi(\Delta B+\Delta B_\text{v})/\Delta B= 2\pi + \Delta\varphi_\text{v} \Delta N$. Finally we get the formula $\Delta N=(2\pi/\Delta\varphi_\text{v})(\Delta B_\text{v}/\Delta B)$, which defines the average number of vortices entering the electrodes per one period of the HV-oscillation. Thus, for example of sample N7, the number of entering vortices, per one period, is $\Delta N=2$. In this estimate we have used $\Delta B_\text{v}=0.078$ G, $\Delta B=0.2$ G, and $\Delta\varphi_\text{v}=1.22$ (to be discussed in the following paragraph). Now we are ready to make an estimate of the average number of vortices entering the electrodes as the applied field is changed by one Gauss, which is $dN/dB=\Delta N/(\Delta B+\Delta B_\text{v})=7.2$, assuming that $N(B)$, the number of vortices versus magnetic field, is linear. 

The estimate above required us to make an assumption that each vortex generates a phase bias $\Delta\varphi_\text{v}=1.22$, on average. To justify this, consider one vortex in the center of one of one of the electrode, illustrated Fig.~\hyperref[fig:setup_d]{\ref{fig:setup_d}}. Then, according to Ref.~\cite{Golod2010PRL}, $\Delta\varphi_\text{v}$ is equal to the polar angle $\Theta_\text{v}$ subtended by the line connecting the entrance points of the two bridges leading to the JJs forming the SQUID. Therefore, for our samples, $\Delta\varphi_\text{v}=\Theta_\text{v} = 2\tan^{-1}[2Z/(25-Y)]$ [$Y$ and $Z$ are shown in Fig.~\hyperref[fig:setup]{\ref{fig:setup_d}}]. Therefore, for Sample N7, the phase bias imposed by one vortex located in the center of the electrode is $\Delta\varphi_\text{v}=1.22$ rad. The geometry of the electrodes of the sample N1 is very similar to N7, therefore we will assume the same phase bias per vortex for the sample N1.

Of course, the number of vortices entering the electrodes exhibits fluctuations. In what follows we present a different approach to analyze and average out these fluctuations. First, we calculate the difference between the consecutive sweet spot fields $\Delta B_n$=$B_n-B_{n-1}$. The result is plotted in Fig.~\hyperref[fig:vortexcount]{\ref{fig:vortexcount_b}}. With these notations $\Delta B_n-\Delta B$ is the increase of the measured period above the vortex-free period $\Delta B$. The number of vortices entering the electrodes between two adjacent sweet spots, $\Delta N_n$, can be estimated using formula $\Delta N_n/(\Delta B_n-\Delta B)$=$\Delta N /\Delta B_\text{v}$ because $\Delta N =<\Delta N_n>$ and $\Delta B_\text{v}=<\Delta B_n-\Delta B>$. Here it is assumed that the number of vortices entering per period is linearly proportional to the period. Thus computed number $\Delta N$ is shown on the right axis in Fig.~\hyperref[fig:vortexcount]{\ref{fig:vortexcount_b}}. Finally, we integrate $\Delta N_{n}$ with respect to $n$ to output the total number of vortices, $N_n$, versus the sweet spot index $n$. The result is shown in Fig.~\hyperref[fig:vortexcount]{\ref{fig:vortexcount_c}}. From this plot one can estimate that that 38 vortices enter as the index is increased by 20 (remember that $\Delta n=1$ corresponds to one period). Thus one obtains 1.9 vortices per period. Since the period equals 0.278 G, on average, one estimates that $dN/dB=$6.8 vortex/G, in the case of sample N7. Since N1 has about the same size of the electrodes as N7, we use the same conversion factor $dN/dB$ for N1.

Now we discuss our third approach to estimate the vortex entrance rate $dN/dB=(N_{n+1}-N_n)/\Delta B$. This approach is based on the observation that the function $N_n$, computed by the algorithm outlined above, exhibits a stepwise increase as the magnetic field is swept linearly [Fig.~\hyperref[fig:vortexcount]{\ref{fig:vortexcount_c}}]. The steps are made more noticeable by placing the horizontal dashed lines. The spacing between the lines is constant and they serve as guide to the eye. The step size turns out  almost constant. We speculate that each step corresponds to the entrance of a single vortex which is effectively coupled to the qubit. This scenario assumes that not all vortices present in the loop are sufficiently well coupled to the supercurrent generated by the qubit but only those which enter the area near the SQUID loop. The physical reason for this is the fact that the current tends to be concentrated near the edges due to the Meissner effect. At the same time, many other vortices get pushed in the middle of the electrode, thus making their impact on the qubit very minimal. It is naturally expected that vortices entering the electrodes near the loop would make a relatively large impact on the change of the period, $B_{n+1}-B_{n}$ and therefore can cause an sharp increase in the estimated $N_n$ number. Thus the steps apparent in Fig.~\hyperref[fig:vortexcount]{\ref{fig:vortexcount_c}} represent the vortices effectively coupled to the qubit, and only these vortices are relevant for our estimate of the relaxation rate. In this scenario, the total number of effectively coupled vortices equals the number of steps, i.e. equals 6. These 6 vortices have entered over the interval of 5.6 Gauss. Thus the effective entrance rate can be taken as $dN/dB$=6/5.6=1.07 vortex/G. This estimate provides the lower bound for $dN/dB$.

Remember that the experimental relaxation rates per Gauss $d\Gamma/dB$, obtained from Fig.~\hyperref[fig:T1rate]{\ref{fig:T1rate}}, are 78.5 and 43.7 kHz/G for samples N1 and N7. Now these values need to be divided by 6.8, which is the rate of the vortex entrance, $dN/dB$. Thus we conclude that the experimental relaxation rate is $\Gamma_\text{v}=11.5$ and $\Gamma_\text{v}=6.4$ kHz/vortex, for samples N1 and N7, calculated using $\Gamma_\text{v}=(d\Gamma/dB)/(dN/dB)$. These values are somewhat smaller than the theoretical estimates of 89 kHz and 48 kHz/vortex. This fact serves as indirect evidence that the number of vortices is overestimated.  

With the conversion factor obtained by our third method, based on the observation of the steps, the experimental relaxation rates per vortex become 73 and 41 kHz/vortex for sample N1 and N7. These values are in good agreement with the calculated values, 89 and 48 kHz/vortex, for samples N1 and N7. Thus the approach based on the step counting (i.e., our third method) appears to be the most accurate for the estimation of the effective number of vortices effectively coupled to the qubit. 

\bibliography{MeissnerQubitPaper}

\begin{thebibliography}{39}
\bibitem [{Qfa()}]{Qfactor}%
  \BibitemOpen
  \href@noop {} {}\bibinfo {note} {$Q_\text{e}$ was set by adjusting the length
  of the center pins of SMA connectors in the cavity. $Q_\text{e}$ can be
  calculated, given measured interal quality factor ($Q_\text{i}$), loaded
  quality factor ($Q_\text{L}$) and insertion loss (IL) in the following
  expression: $1/Q_\text{L} = 1/Q_\text{e} + 1/Q_\text{i}$,
  $IL(\textnormal{dB}) =
  20\log[Q_\text{i}/(Q_\text{i}+Q_\text{e})]$.}\BibitemShut {Stop}%
\bibitem [{noi()}]{noiseratio}%
  \BibitemOpen
  \href@noop {} {}\bibinfo {note} {Although the smaller input coupling
  increases the signal-to-noise ratio, it should not be too small since
  otherwise the spectroscopic power at the input port of the cavity required to
  drive a qubit would be unattainably high beyond the possible output power of
  a microwave source.}\BibitemShut {Stop}%
\bibitem [{bar()}]{barefreq}%
  \BibitemOpen
  \href@noop {} {}\bibinfo {note} {Tha bare cavity frequency $f_\text{c}$ is
  the resonant frequency of the cavity without dipersive shift. It is the
  cavity frequency at large drive power where the cavity reaches the state of a
  near-unity transmission. $f_\text{c}$ does not depend on the qubit transition
  frequency and state.}\BibitemShut {Stop}%
\end{thebibliography}

\end{document}